\documentclass[twocolumn,showpacs,preprintnumbers,amsmath,amssymb]{revtex4-1}

\usepackage{graphicx}
\usepackage{dcolumn}
\usepackage{bm}
\usepackage[cp1250]{inputenc}
\usepackage{ifthen}

\begin{document}
\newcommand{\bk}{\mathbf{k}}
\newcommand{\bQ}{\pmb{Q}}
\newcommand{\mb}{\mathbf}
\newcommand{\ua}{\uparrow}
\newcommand{\da}{\downarrow}
\newcommand{\dg}{^\dagger}
\newcommand{\ef}{\epsilon_f}
\newcommand{\os}{\overline{\sigma}}
\newcommand{\eq}{\begin{equation}}
\newcommand{\eqx}{\end{equation}}
\newcommand{\eqn}{\begin{eqnarray}}
\newcommand{\eqnx}{\end{eqnarray}}

\title{Theory of Unconventional Superconductivity in Strongly Correlated Systems:\\
 Real Space Pairing and Statistically Consistent Mean-Field Theory \\
 - in Perspective\footnote{This paper represents a detailed elaboration of the plenary talks delivered at \emph{the European Conference on Magnetism} in Pozna\'{n} (2011), at \emph{Moscow International Symposium on Magnetism} (2011), and at \emph{the National School on Superconductivity} in Kazimierz Dolny (2011).}}


\author{J\'ozef Spa\l{}ek}
\email{ufspalek@if.uj.edu.pl}
\affiliation{Marian Smoluchowski Institute of Physics, Jagiellonian University, Reymonta~4, 30-059 Krakw, Poland;\\
Faculty of Physics and Applied Computer Science,~AGH University of Science and Technology, Reymonta 19, 30-059 Krakw, Poland}

\date{\today}

\begin{abstract}
In this brief overview we discuss the principal features of \emph{real space pairing} as expressed via corresponding low-energy (\emph{t-J} or periodic Anderson-Kondo) effective Hamiltonian, as well as consider concrete properties of those unconventional superconductors. We also rise the basic question of \emph{statistical consistency} within the so-called \emph{renormalized mean-field theory}. In particular, we provide the phase diagrams encompassing the stable magnetic and superconducting states. We interpret real space pairing as correlated motion of fermion pair coupled by short-range exchange interaction of magnitude $J$ comparable to the particle renormalized band energy $\sim tx$, where $x$ is the carrier number per site. We also discuss briefly the difference between the real-space and the paramagnon - mediated sources of superconductivity. The paper concentrates both on recent novel results obtained in our research group, as well as puts the theoretical concepts in a
  conceptual as well as historical perspective. No slave-bosons are required to formulate the present approach.
\end{abstract}

\pacs{71.27.+a, 74.70.Tx, , 74.20.-z}

\maketitle

\section{Introduction: Appearance of strongly correlated states and pairing by exchange interactions}

The analysis of the unconventional (non-BCS) superconductivity is essentially limited to that appearing  in the correlated fermion systems. We term the system \emph{correlated} if the interaction magnitude  between  the particles $\overline{V}$ is comparable or larger to their single-particle (kinetic, band) energy $\overline{E}_{B}$. In the extreme situation when $|\overline{E}_{B}|\ll \overline{V}$, we speak about the strongly correlated systems. Obviously, this simple theoretical criterion must be translated into the specific experimental features distinguishing those systems from other normal metallic, magnetic, and superconducting systems.

First of all, the short-range strong repulsive Coulomb interaction (as compared to the so-called bare single-particle energy) will hamper the individual-particle motion, and hence increase largely its effective mass $m^{*}$, which in turn, will show up in the strongly enhanced value of the linear-specific heat coefficient $\gamma \sim m^{*}$. Also, the strong electron-electron interaction leads to the corresponding temperature ($T$) dependence of the resistivity $\rho(T)-\rho(0) \equiv AT^{2}$, since the coefficient $A \sim (m^{*})^{2}$. All of these features appear already in \emph{the Landau Fermi-liquid theory}, together with an explanation of enhancement of the Pauli susceptibility $\chi\sim m^{*}$ and in addition, with the appearance of collective sound-like excitations.

The situation has changed decisively with the observation of a singular behavior  of $\gamma,\; \chi$, and $A$, but with the ratio $\gamma/\chi$ remaining finite, which appear near the metal-insulator (\emph{Mott-Hubbard}) transition. The transition is associated closely with localization of correlated carriers and is signaled additionally by the transition from \emph{the Slater}- to \emph{the Heisenberg-type of antiferromagnetic ordering}. In such manner, the divergences on the metallic side define the borderline of the metallic-state stability. The transition occurs for specific (odd) number of electrons, e.g. for a half-filled-band configuration of starting electrons, and thus those systems can be easily distinguished from either the band or \emph{the Kondo-type insulators}, for which the fully-occupied bands have even number of electrons, separated in each case by a gap from empty (conduction) states. Detailed studies of \emph{the Mott-Hubbard systems} near the metal-insulator transitions were carried out in the seventies through the nineties of the XX century \cite{a1}. They were subsequently supplemented with  detailed studies of orbital, charge, and stripe orderings in various Mott insulators, i.e. the Mott-Hubbard systems on the insulating side.

The Mott-Hubbard metal-insulator transitions are usually discontinuous, with isolated critical point on the higher-temperature side (\emph{the classical critical point}) and a possible quantum critical point at the antiferromagnetic-paramagnetic boundary \cite{a1}. In a simple modeling of this transitions, they are driven by a competition between the single particle energy, as represented by band energy $\overline{E}_{B}<0$ (per particle, effectively characterized, by bare bandwidth $W$ of fermions)  and the repulsive Coulomb energy $\overline{V}>0$ (represented by the magnitude $U$ of the intraatomic Coulomb interaction). The physics, for given number of electrons $n$ per active atomic site, is characterized then by changing $U/W$ ratio (or effectively, by exerting external pressure  which reduces the $U/W$ ratio). The situation is different when we have an orbitally degenerate system, still with one electron per active band per site. For example, for $n=1$ the corresponding Mott insulator may take the form of a ferromagnetic insulator with an antiferromagnetic orbital ordering \cite{a2} or the state with orbital-selective metal-insulator transition.

A basic question arises what happens if we vary the electron concentration (the band filling) instead of changing $U/W$ ratio for given $n$. This situation is quite distinct from that when changing $U/W$. This is because in the situation with partial (non-half) filling the metallic state is stable even in the strong-correlation limit $U/W \gg 1$, if only the disorder effects associated with e.g. intentional doping, do not induce the carrier localization of holes or electrons in a weakly \emph{doped Mott insulator} \cite{a3}. The doped Mott insulators and the heavy-fermion systems are exactly the systems of that type. While the high-$T_C$ cuprates such as $La_{2-x}Sr_{x}CuO_{4}$ can be represented as doped Mott insulator with concentration $x \lesssim 0.3$ of holes per $CuO_{2}^{2-}$ active unit, the cerium heavy-fermion stoichiometric compounds such as $CeAl_{3}$ or $CeCoIn_{5}$ can be regarded as \emph{almost localized systems} with $\delta \lesssim 0.05$ holes in nominally
 $Ce^{3+}$ $4f^{1}$ electronic configuration (i.e. the $Ce$ valence is $3+\delta$). In both systems the $3d$ (for the cuprates) and the hybridized $4f-5d-6s$ (for the cerium compounds) strongly correlated electrons are itinerant, which may transform to the localized states under a moderate change of stoichiometry, pressure or applied magnetic field. For both systems we assume that $U/W \gg 1$ and $x\equiv 1-n\ll 1$.

One may ask whether the clear borderline for the change of behavior, corresponding to metal-insulator transition, survives also for the doped systems. In other words, whether there is a clear distinction between the limit of weakly or moderately correlated fermions from one side and the regime of strong correlations for $n\neq 1$ from the other, when no metal-insulator transition can occur. The general assumption usually made is that such a dividing line indeed exists (albeit of a crossover type) and this statement represents one of the fundamental hypotheses of \emph{the theory of strongly correlated systems}, even though its properties have not been proved convincingly. In the case of the cuprates such a line can be drawn on the temperature $T$-doping $x$ plane and terminate at the middle point $x=x_{c1}\sim 0.15$, of the superconducting dome and separates a some sort of Fermi liquid extends from that line to its upper-end concentration \cite{a3,a4} $x=x_{c2}\sim 0.3$. In t
 he case of heavy fermion systems, the existence of such line is suggested through the appearance of antiferromagnetism or \emph{metamagnetism} in a strong applied magnetic field or through the transition to the fluctuating (mixed) valence state as a function of pressure or else, to the localized moment regime with the increasing temperature.

In our brief overview we assume that such (crossover) line exists and thus limit ourselves to the analysis of normal and superconducting states in the limit of strong correlations. This limiting regime is defined as the one, in which the probability of double site  (orbital) occupancy, $d^{2}\equiv\langle n_{i\uparrow}n_{i\downarrow\rangle}$ is vanishingly small (so formally, $d^{2}=0$). This is also the limit, where we cannot start from the Hartree-Fock representation of the electronic states as the reference state for the further analysis. Instead, we will start from \emph{the so-called statistically-consistent Gutzwiller} or \emph{Fukushima mean-field approach}, devised in our group in the last two years \cite{a5}. We regard this approach as \emph{the first consistent renormalized mean-field approach} of strongly correlated fermionic systems, here employed to the description of superconducting state with real space pairing.

One important point should be raised. Namely, the pairing interaction is of real-space type and is driven by \emph{the kinetic exchange interaction} under the name of \emph{the t-J model} for a single narrow-band case and \emph{the Anderson-Kondo model} (or \emph{the t-J-V model}) with the hybrid pairing for a two-band for a two-band situation. Both of these models were introduced by the present author some time ago in the context of magnetism and normal correlated states and later extended to the description of superconductivity in those systems  \cite{a6, a7}. A somewhat more detailed personal account, of the transformation to the corresponding effective Hamiltonians into forms expressing explicitly the superconducting pairing, is provided in Appendix A.

The paper is organized as follows. In the next Section we discuss the universal aspects of real space pairing, namely the fact it is induced by a correlated motion of pair of particles throughout the lattice. In Section III we summarize the principal features of the so-called Renormalized Mean-Field Theory (RMFT) together with the consistency conditions. Particular emphasis is put on the phase diagram encompassing the unconventional superconducting states. In Section IV we overview selected physical properties of the superconducting state and its coexistence with magnetism. In such IV we put into a perspective the whole approach. Appendices A-E are to provide details of the approach and to put it on a firmer formal grounds.

\section{$\pmb{t}$\emph{-J} and Anderson-Kondo lattice models: real space pairing}

In this Section we overview briefly the universal character of real space pairing, i.e. its applicability to both high-$T_{C}$ and heavy-fermion systems. By \emph{the real space pairing} we understand a correlated electron-pair motion (hopping) of partners which are coupled by rather strong exchange interaction (i.e. comparable on the scale of the Fermi energy). The simplest model of one-band strongly correlated metallic systems is the Hubbard model, whereas the Anderson-lattice model reflects the corresponding two-band situation when the starting atomic (say $4f$) electrons are hybridized with the uncorrelated band ($5d-6s$) conduction electrons. Some time ago \cite{a6,a7} both of these models were transformed out by us to the form  expressing explicitly the strong correlations, on the low-energy (thermodynamic) scale, by projecting out in a \emph{precise manner} the doubly occupied atomic configurations and in this manner reducing the effect of the strong intraatomic Coulom b repulsion by replacing them with dynamic effects on the low-energy scale. Below we summarize briefly each of the two above models.

\subsection{\emph{t-J} model and its extension}

Let us start from the extended Hubbard Hamiltonian for a single narrow band of $s$-type:
\begin{equation}
\mathcal{H}=\sum_{ij\sigma}\!^{'}t_{i j}a_{i\sigma}^{\dagger}a_{j\sigma}+
U\sum_{i}n_{i\uparrow}n_{i\downarrow}+\frac{1}{2}\sum_{ij}\!^{'}K_{ij}n_{i}n_{j},
\label{1}
\end{equation}
where the primed summations mean that we take only $i\neq j$ terms in both the hopping (the first) and the intersite Coulomb (the third) terms. The second term is the celebrated Hubbard term representing the energy increase when the site (Wannier) state is doubly occupied. The principal feature here is that the Coulomb repulsive interactions are developed systematically and express respectively the interaction between two electrons on the same site (with magnitude $U$) which represents (by far) the largest energy scale, as well as between the electrons on neighboring sites $i\neq j$, with magnitude $K_{ij}$ and neglect all more distant, interactions, since we assume that Wannier states are strongly localized ( \emph{the tight-binding approximation}). In other words, this atomic representation differs drastically from the concept of a lattice electron gas as a starting point, since the long-range nature of the repulsive Coulomb interaction is cut off. The decisive step was taken \cite{a6} to derive an effective Hamiltonian out of (\ref{1}) expressing low-energy dynamics, i.e. which contains high-energy processes as virtual transitions in the second order order. The resultant effective Hamiltonian in the lowest Hubbard band is:
\begin{eqnarray}
\widetilde{\mathcal{H}}
&=&\sum_{ij\sigma}\!^{'}t_{ij}b_{i\sigma}^{\dagger}b_{j\sigma}+
\sum_{ij\sigma}\!^{'}\frac{2t_{ij}^{2}}{U-K_{ij}}\left(\pmb{S}_{i}\cdot \pmb{S}_{j}-\frac{1}{4}\nu_{i}\nu_{j}\right) \nonumber \\
&+&\sum_{ijk\sigma}\!^{''}\frac{t_{ij}t_{jk}}{U-K_{ij}}
\left(b_{i\sigma}^{\dagger}\nu_{j\bar{\sigma}}b_{k\sigma}-
b_{i\sigma}^{\dagger}S_{j}^{\bar{\sigma}}b_{k\bar{\sigma}}\right)\nonumber \\
&+&\frac{1}{2}\sum_{ij}\!^{'}K_{ij}\nu_{i}\nu_{j}.
\label{2}
\end{eqnarray}
The double-primed summation means $i\neq j\neq k \neq i$. The projected fermion operators are:
$b_{i\sigma}^{\dagger} \equiv a_{i\sigma}^{\dagger}\left(1-n_{i\bar{\sigma}}\right)$,
$b_{j\sigma} \equiv a_{j\sigma}\left(1-n_{j\bar{\sigma}}\right)$,
$\nu_{i\sigma}\equiv b_{i\sigma}^{\dagger}b_{i\sigma}$,
$\nu_{i}=\sum_{\sigma}\nu_{i\sigma}$, and
$\pmb{S}_{i}\equiv\left(S_{i}^{\sigma},S_{i}^{z}\right)\equiv
\left(b_{i\sigma}^{\dagger}b_{i\bar{\sigma}},
\left(\nu_{i\uparrow}-\nu_{i\downarrow}\right)/2\right)$.
We see that apart from a restricted (projected) hopping,
with no double occupancies present,
we have also present an antiferromagnetic kinetic exchange interaction with exchange integral $J_{ij}\equiv 4t_{ij}^{2}/(U-K_{ij})$ \cite{a8} (the second term), here generalized to the case of itinerant fermions, since the spin operators are explicitly expressed in terms of fermionic operators. The third term describes three-site hopping processes, and the last - the residual repulsive intersite Coulomb interaction. The dynamical processes in the second order, taken into account in (\ref{2}) are shown in Fig 1. We also included there the renormalized
single-particle hopping.
\begin{figure}
\includegraphics[width=0.35\textwidth]{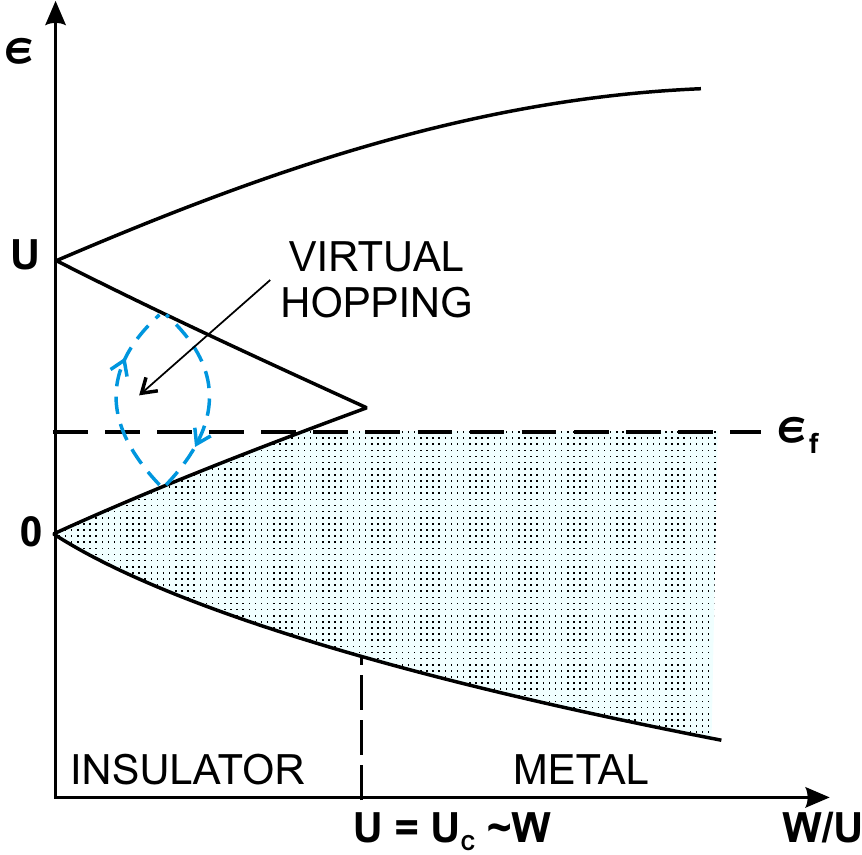}
\vspace{1cm}{
\includegraphics[width=0.4\textwidth]{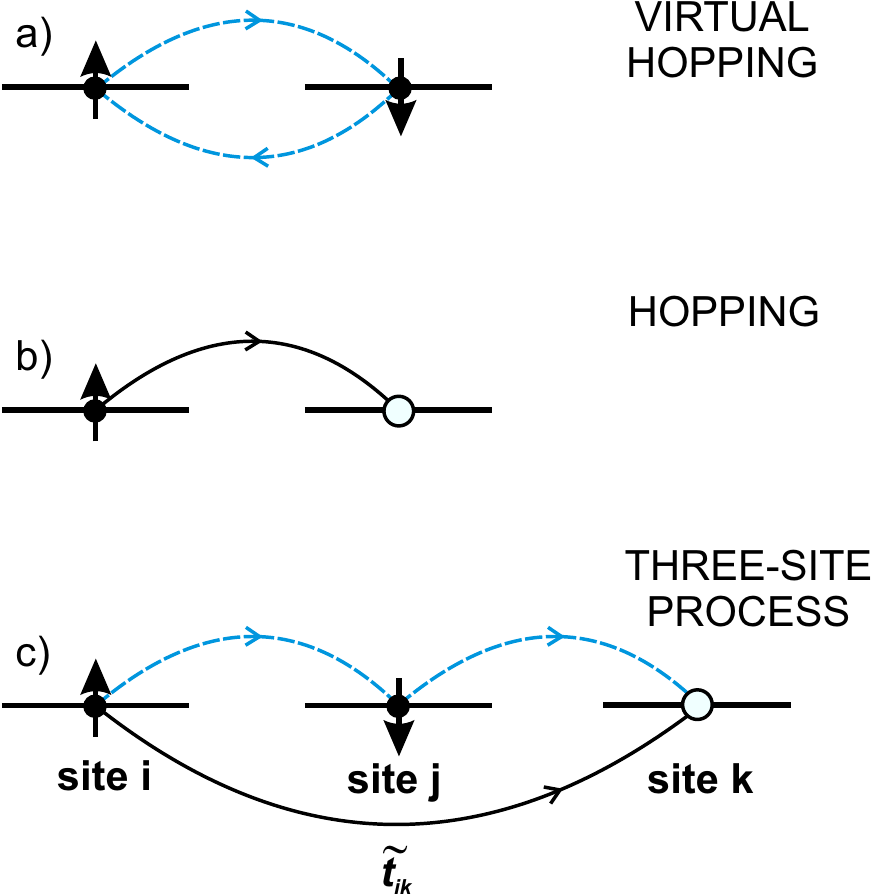}}
\caption{(Color online) Top: schematic representation of the Hubbard subbands. The virtual hopping processes correspond to the virtual transitions lower $\rightarrow$ upper $\rightarrow$ lower Hubbard subbands (process $\left.\mbox{a}\right.$). Bottom: three possible intersite hopping processes (in real space) determining the dynamics of Hamiltonian (2) in the lower Hubbard subband (i.e. for the band filling $n\leq 1$): $\left.\mbox{a}\right)$ virtual hopping between single occupied neighboring sites that leads to their antiferromagnetic (kinetic) exchange interaction; $\left.\mbox{b}\right)$ single-particle hopping between the single occupied and the empty sites; and $\left.\mbox{c}\right)$ three-site hopping between the singly occupied and the empty site via an intermediate single occupied site with opposite spin.}
\end{figure}

One has to mention that the projected fermion operators $\{b_{i\sigma}\}$ and $\{b_{i\sigma}^{\dagger}\}$ do not obey the usual fermion anticommutation relation, as we have that
\begin{eqnarray}
\{b_{i\sigma},b_{j\sigma'}^{\dagger}\}&=&\delta_{ij}\left[\left(1-n_{i\bar{\sigma}}\right) \delta_{\sigma\sigma'}+S_{i}^{\bar{\sigma}}\left(1-\delta_{\sigma\sigma'}\right) \right] \nonumber \\
\{b_{i\sigma},b_{j\sigma'}\}&=&0
\label{3}
\end{eqnarray}
This last property creates a basic formal complication, as we are working now with projected fermion operators. We can say that the effective model represents, in principle a fermionic quantum liquid, \emph{which is not a Landau Fermi liquid} (usually termed imprecisely as: \emph{non-Fermi liquid}). As a first approximation, we discuss in the next Section an effective (\emph{almost localized}) Fermi liquid and demonstrate the usefulness of the approach as a renormalized mean field theory. Note that the Hamiltonian (\ref{2}), is valid for the band filling $n\leq 1$. It can be easily extended to the situation when $n\geq 1$ by employing the hole language.

The model (\ref{2}), mainly in the limit of $K_{ij}\equiv 0$, was used to discuss the magnetic, charge-density, and mixed (stripe) phases. However, a new impetus to study the model was the idea [3], that the \emph{t-J} model (i.e. Hamiltonian (\ref{2}) with $K_{ij}\equiv 0$) can be used to describe the so-called \emph{real space pairing}. The real-space pairing operators may be defined in the following manner
\begin{equation}
\left\{ \begin{array}{cl}
B_{ij}^{\dagger}\equiv\frac{1}{\sqrt{2}}\left(b_{i\uparrow}^{\dagger} b_{j\downarrow}^{\dagger}-b_{i\downarrow}^{\dagger}b_{j\uparrow}^{\dagger}\right),\\
B_{ij}\equiv \frac{1}{\sqrt{2}}\left(b_{i\uparrow} b_{j\uparrow}-b_{i\uparrow}b_{j\downarrow}\right).
 \end{array}\right.
\label{4}
\end{equation}
In this manner, both two- and three-site terms can be recast to the closed form and the effective Hamiltonian transforms to the form
\begin{eqnarray}
\bar{\mathcal{H}}&=&\sum_{ij\sigma}\!^{'}t_{ij}b_{i\sigma}^{\dagger}b_{j\sigma}-
\sum_{ijk}\!^{'}\frac{2t_{ij}^{2}}{U-K_{ij}}
B_{ij}^{\dagger}B_{kj} \nonumber \\
&+&\frac{1}{2}\sum_{ij}\!^{'}K_{ij}\nu_{i}\nu_{j},
\label{5}
\end{eqnarray}
where now both the terms with $k=i$ and $k\neq i$ are incorporated into the second term. The expressions (\ref{4}) represent the projected spin-singlet creation and annihilation operators located on the pair of sites $(i,j)$ with $i\neq j$, since $B_{ii}^{\dagger}=B_{ii}\equiv 0$. Hence, no ionic mixture to such two-site spin-singlet state appears. Also:
\begin{equation}
B_{ij}^{\dagger}B_{ij}\equiv -\left(\pmb{S}_{i}\cdot \pmb{S}_{j}-\frac{1}{4}\nu_{i}\nu_{j}\right),
\label{6}
\end{equation}
i.e. the "number operator for local singlets" is equivalent to the generalized Dirac exchange operator (r.h.s). In other terms, the description in terms of itinerant-spin interaction language is equivalent to the description in terms of local $\langle i,j \rangle$ itinerant-spin singlet. In the other words, the antiferromagnetic ordering coming from the kinetic exchange interaction and the kinetic-exchange or real-space singlet pairing should be regarded as  equivalent ways of describing the interaction i.e., express a different type of ordering, which should be treated on equal footing. What is even more important, such paired state is directly included by the correlated motion of those singlet pairs, as discussed below.

One general remark is in place here. The operators $B_{ij}^{\dagger}$ and $B_{ij}$ express explicitly the singlet nature of the pairs if $\langle B_{ij}^{\dagger}\rangle\neq 0$. Normally, one operates only with either $\langle c_{i\uparrow}^{\dagger}c_{j\downarrow}^{\dagger}\rangle$ or alternatively, with $\langle c_{i\downarrow}^{\dagger}c_{j\uparrow}^{\dagger}\rangle$ \cite{a16}. But then, one implicitly assumes that $\langle c_{i\uparrow}^{\dagger}c_{j\downarrow}^{\dagger}\rangle=-\langle c_{i\downarrow}^{\dagger}c_{j\uparrow}^{\dagger}\rangle$. Strictly speaking, this relation should be checked out explicitly, e.g. evaluating those two averages separately. Nota bene, this explicit checkout helps in distinguishing between the singlet pairing and the triplet pairing (both components with $S^{z}=0$). Namely, in the latter case $\langle c_{i\uparrow}^{\dagger}c_{j\uparrow}^{\dagger}\rangle=+\langle c_{i\downarrow}^{\dagger}c_{j\downarrow}^{\dagger}\rangle$. The spin nature of
  real-space pairing is particularly obscured, if we consider the coexistence between magnetism and superconductivity, as discussed below.

Additional question is concerned with the presence of the repulsive term $\sim K_{ij}$, which is usually neglected in the analysis of high-temperature superconductivity within the \emph{t-J} model. Omission of this term is justified by the circumstance that is regarded as contributing only the reference energy, i.e. is the same for the phases under consideration. It may be also regarded as compensated by the electron-lattice interaction which leads  effectively to $K_{ij}<0$ \cite{a17}. If this would be the case, then that the pairing part induced by the case  $J_{ij}$ does not lead to isotope effect, whereas the part $\sim K_{ij}<0$ does. Since the isotope effect in high-$T_{c}$ system is small, it means that the kinetic exchange part of pairing in dominant.

\subsection{Hybrid (Kondo-type) pairing in Anderson-lattice model}

The Anderson-lattice model represents the simplest two-orbital model with a coherent mixture (hybridization) of strongly correlated and uncorrelated electrons. It is the canonical model of heavy-fermion system if the orbital degeneracy of e.g. $4f^{1}$ electronic configuration of $Ce^{3+}$ ions is not crucial (e.g.the lowest double $\Gamma_{7}$ is the only important crystal-field level which hybridizes with conduction electrons originating from the $5d-6s$ itinerant states). The starting Hamiltonian has the following form in the site (Wannier) representation:
\begin{eqnarray}
\mathcal{H}=\sum_{mn\sigma}\!^{'}t_{mn}c_{m\sigma}^{\dagger}c_{n\sigma}+
\epsilon_{f}\sum_{i\sigma}N_{i\sigma}+U\sum_{i}N_{i\uparrow}N_{i\downarrow} \nonumber\\
+\sum_{im\sigma}\left(V_{im}a_{i\sigma}^{\dagger}c_{m\sigma}+V_{im}^{*}
c_{m\sigma}^{\dagger}a_{i\sigma}\right)+U_{fc}\sum_{im}N_{i}n_{m},
\label{7}
\end{eqnarray}
where $(i,j)$ label starting atomic $(a)$ states, $(m,n)$ label starting delocalized (conduction, $c$) states, and $N_{i\sigma}\equiv a_{i\sigma}^{\dagger}a_{i\sigma}$. $V_{im}$ represents the hybridization matrix element. The strongly-correlated aspect shows up through the circumstance that both $|V_{im}|$ and $|t_{mn}|\ll U$, but the position of the atomic level $\epsilon_{f}\sim V_{im}$, so not the whole hybridization term can be transformed out via the Schrieffer-Wolff-type of transition to the Kondo lattice type of model \cite{a7, a19}.

To adopt the model to the situation with strong correlations, one performs the transformation, in which as in the narrow-band case, the double occupancies of atomic states are excluded and replaced by the virtual processes leading among others to the Kondo interactions between the electrons of the two subsystems \cite{a7}. Explicitly, the full effective Hamiltonian in the second order in $V/U$ and representing the low-energy dynamics, takes the form:
\begin{eqnarray}
\widetilde{\mathcal{H}} &=& \sum_{mn\sigma}\!^{'}\left[t_{mn}c_{m\sigma}^{\dagger}c_{n\sigma}-
\sum_{i}\frac{V_{im}^{*}V_{mi}}{U+\epsilon_{f}}\nu_{i\bar{\sigma}}
c_{m\sigma}^{\dagger}c_{n\sigma}\right]\nonumber \\
&+&
\sum_{i\sigma}\epsilon_{f}\nu_{i\sigma}
+\sum_{im\sigma}V_{im}\left(b_{i\sigma}^{\dagger}c_{m\sigma}+H.c.\right)\nonumber \\
&+&
\sum_{im}\frac{2|V_{im}^{*}|^{2}}{U+\epsilon_{f}}
\left(\pmb{S}_{i}\cdot\pmb{s}_{m}-\frac{1}{4}\nu_{i}n_{m}\right)  \label{8}\\
&+&
\frac{1}{U+\epsilon_{f}}\sum_{imn\sigma}V_{mi}V_{in}^{*}\left(S_{i}^{\bar{\sigma}}
c_{m\bar{\sigma}}^{\dagger}c_{n\sigma}+\nu_{i\bar{\sigma}} c_{m\sigma}^{\dagger}c_{n\sigma}\right),\nonumber
\end{eqnarray}
where $\nu_{i\sigma}\equiv b_{i\sigma}^{\dagger}b_{i\sigma}\equiv a_{i\sigma}^{\dagger}a_{i\sigma}\left(1-N_{i\bar{\sigma}}\right)$, and $\pmb{s}_{m}\equiv \left(s_{m}^{\sigma},s_{m}^{z}\right)\equiv\left(c_{m\sigma}^{\dagger}c_{m\bar{\sigma}},
(n_{m\uparrow}-n_{m\downarrow})/2\right)$.

The important point to note is that in this Hamiltonian we have both antiferromagnetic Kondo coupling with the exchange constant $J_{im}\equiv 2|V_{im}|^{2}/(U+\epsilon_{f})$ and the residual hybridization in the projected subspace, $\sim (V_{im}b_{i\sigma}^{\dagger}c_{m\sigma}+H.c.).$ In this manner, the itineracy of $f$ electrons is explicitly expressed unless there is a phase transition to the localized Mott state for $f$ electrons (i.e. to the state with $\langle\nu_{i}\rangle=1$, see below). The dynamic (virtual) processes in the second order and taken into account in the effective Hamiltonian (\ref{8}), are displayed in Fig.2. There, we also show explicitly the
remaining processes: hoping in the conduction band and the effective $f$-electron hopping
(the bottom most part).
\begin{figure}
\begin{center}
\includegraphics[width=0.4\textwidth]{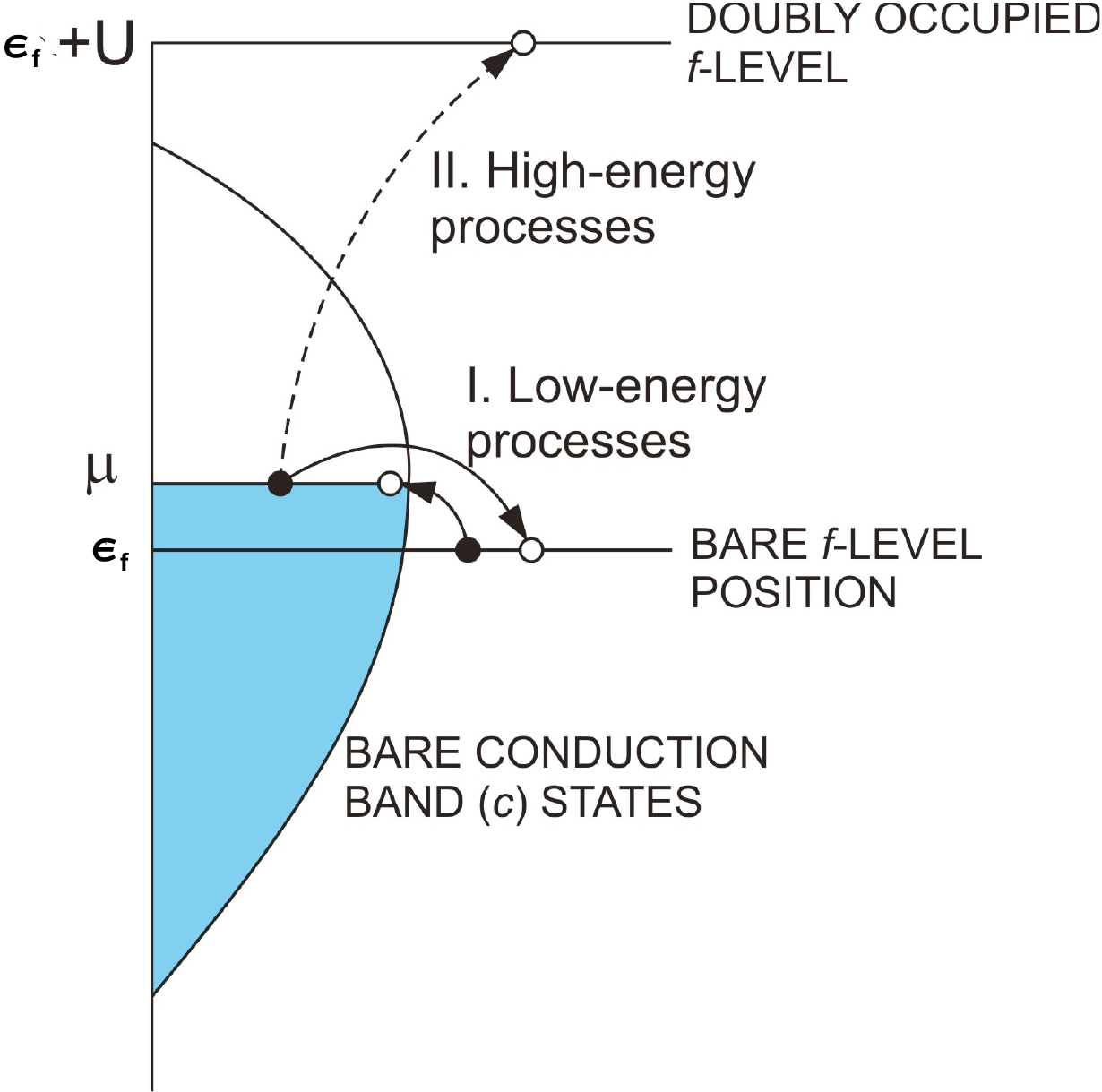}
\vspace{1cm}{
\includegraphics[width=0.4\textwidth]{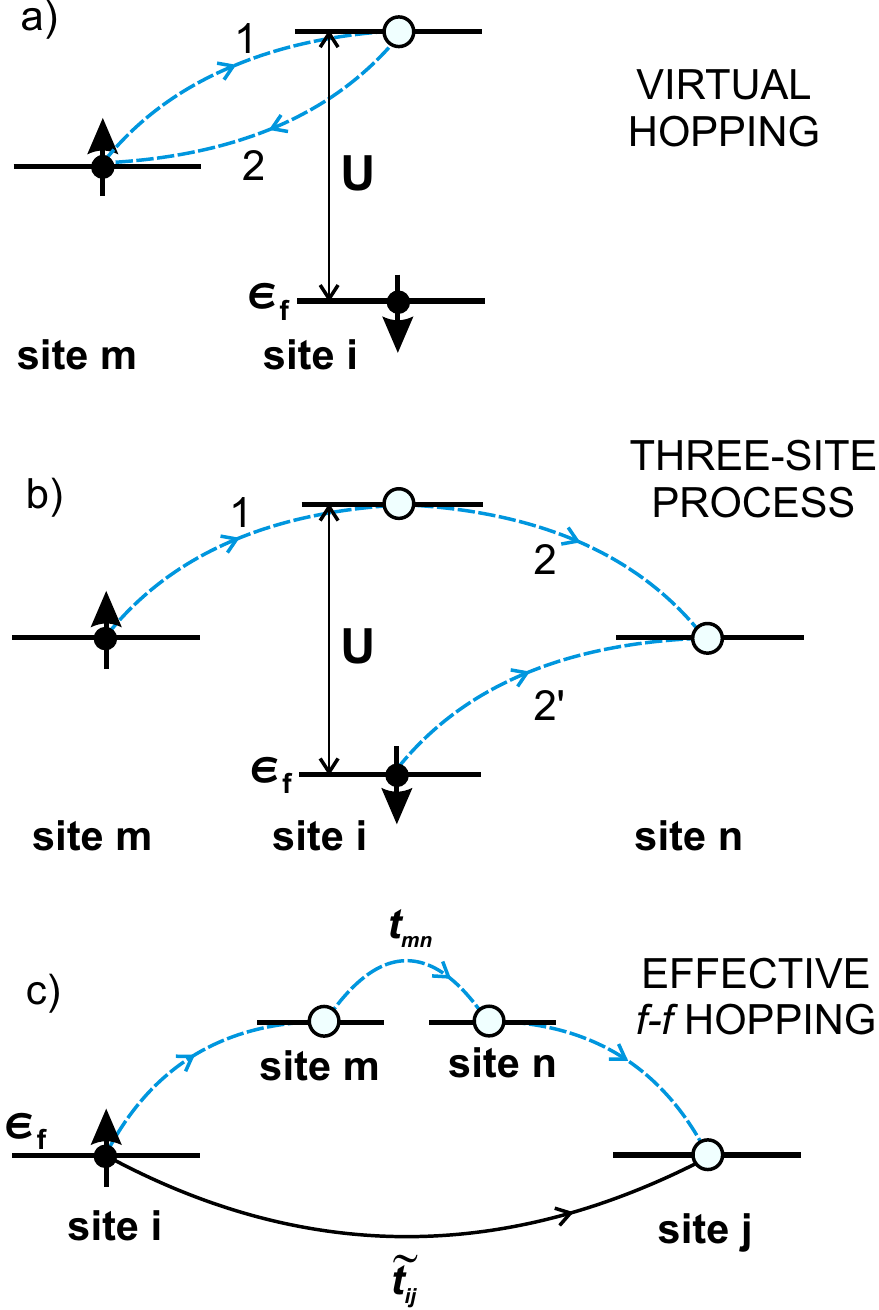}}
\end{center}
\caption{(Color online) Top: schematic representation of hybridization-induced process as f-level occupation dependent hopping process and its division into the low- and the high-energy processes. The former (I) corresponds to the presence of the residual hybridization term in (8); the other (II) leads to the Kondo-type coupling which in turn is expressed as a real-space hybrid pairing in (11) in the second order in $V/\epsilon_{f}$. Bottom: the possible interband hopping process in real space involving virtual hopping process to the double occupied $f$-level $\left.\mbox{a}\right)$ and the three-state (site) hopping process ($\left.\mbox{b}\right)$ and $\left.\mbox{c}\right)$, 2 and 2' alternative steps). The arrow (labels) by $\widetilde{t}_{ij}$ represents the resulting intersite $f-f$ hopping process. }
\end{figure}
The second important point is, that in analogy to the real-space spin-singlet pair operators (\ref{5}), we can introduce \emph{the hybrid (Kondo-type)} pair operators, namely define:
\begin{equation}
\left\{ \begin{array}{cl}
b_{im}^{\dagger}\equiv\frac{1}{\sqrt{2}}\left(b_{i\uparrow}^{\dagger} c_{m\downarrow}^{\dagger}-b_{i\downarrow}^{\dagger}c_{m\uparrow}^{\dagger}\right),\\
b_{im}\equiv \frac{1}{\sqrt{2}}\left(b_{i\downarrow} c_{m\uparrow}-b_{i\uparrow}c_{m\downarrow}\right),
 \end{array}\right.
\label{9}
\end{equation}
and rewrite the Hamiltonian in a more closed form, at least when $V_{im}=V_{im}^{*}$, and we neglect the second order contribution to the hopping (the second part in the first term of (\ref{7})). We obtain then
\begin{eqnarray}
\widetilde{\mathcal{H}} &=& \sum_{mn\sigma}\!^{'}t_{mn}c_{m\sigma}^{\dagger}c_{n\sigma}+
\epsilon_{f}\sum_{i\sigma}\nu_{i\sigma} \label{10} \\
&+& \sum_{im\sigma}V_{mi}\left(b_{i\sigma}^{\dagger}c_{m\sigma}+
c_{m\sigma}^{\dagger}b_{i\sigma}\right)-
\sum_{imn}\frac{2V_{im}V_{in}}{U+\epsilon_{f}}
b_{im}^{\dagger}b_{in}.
\nonumber
\end{eqnarray}
Note that, in analogy to (\ref{6}) we have now
\begin{equation}
b_{im}^{\dagger}b_{im}=-\left(\pmb{S}_{i}\cdot\pmb{s}_{m}-\frac{1}{4}\nu_{i}n_{m}\right),
\end{equation}
i.e. both the Kondo interaction and the hybrid singlet pairing in real space appear on the same footing. Both processes are characterized by the Kondo exchange integral $J_{im}^{K}=2|V_{im}|^{2}/(U+\epsilon_{f})$.
This is the Anderson-Kondo Hamiltonian capable of describing both the itinerant nature of the heavy-$f$ electrons and their localization, as well as the \emph{hybrid (Kondo)} paired state in the limit of strong correlations among the hybridized fermions. By analogy with \emph{t-J} model (\ref{5}), the one represented by Hamiltonian (\ref{10}) can be called \emph{the t-J-V model}. Also, when going to the higher, fourth order in $V_{im}/(U+\epsilon_{f})$, we can include additionally the $f$-$f$ pairing via the term $\sim V^{4}/(U+\epsilon_{f})^{3}B_{ij}^{\dagger}B_{kj}$. In effect, both the (orbital selective) localization of $f$ electrons, as well as a two gap superconducting state are encompassed as limiting cases of this coherent quantum liquid composed of two sets of hybridized fermions.

To summarize this Section, the formal expressions of the kinetic exchange and of the Kondo interaction through the same type of real space pairing operators illustrates the universality of the pairing in both the Mott-Hubbard and heavy-fermion systems. Obviously, the practical usefulness of the pairing concept is proved only by showing the stability of the corresponding magnetic and superconducting states induced by this unconventional forms of pairing. In the next Section we summarize formulated by us \cite{a5, a19} the statistically-consistent renormalized mean-field theory, as well as present some exemplary results.

\section{Renormalized mean-field theory: statistically consistent approach (RMFT-SCA)}

\subsection{Gutzwiller approximation with statistical-consistency conditions(SGA) for the Hubbard model}

The above Hamiltonians have a complex form, since they contain the projected fermionic operators (i.e. composite fermion operators with non-fermion anticommutation relations). The question is how to diagonalize Hamiltonian containing such operators, at least in an approximate and consistent manner, so one has the confidence of having a well defined mean-field approach. Parenthetically, note that by making the canonical transformation we include a certain class of higher-order dynamical processes automatically. Hence, a relatively simple approximation (of the Hartree-Fock type) on the effective-Hamiltonian level includes, at least partially, those higher-order dynamic processes. Therefore, instead of rigorous solution impossible to attain for many-particle correlated systems with spontaneous symmetry breakdown, we develop consistency check for approximate solution at hand.

Actually, we can do a bit better than just carrying out the Hartree-Fock type decoupling in the effective Hamiltonian. Namely, the name of the game is \emph{the renormalization of the effective Hamiltonian} combined with a subsequent Hartree-Fock decoupling of the many-body parts still remaining. This procedure is not systematic in the same sense, as the canonical perturbation expansion presented in the preceding Section, but it is in our view , in accordance with our physical intuitive insights into the nature of strong correlations.

Probably, the simplest nontrivial approach to the description of a correlated state is \emph{the Gutzwiller approach}, discussed here in its still simpler form of \emph{a Gutzwiller ansatz}. It relies on a variational approach by postulating the corresponding macroscopic ($N$-body) wave function $|\Psi\rangle$ and a subsequent approximate combinational evaluation of the relevant averages appearing in $\langle\Psi|\widetilde{\mathcal{H}}|\Psi\rangle$ and in
$\langle\Psi|\Psi\rangle$. The wave function respects the exclusion of the double occupancies in real space, in an approximate manner though. An altered approximation scheme has been introduced subsequently by Fukushima \cite{a10}  who introduced additional variational parameters, the so-called fugacity factors, which guarantee that expectation values of the number of particles in the uncorrelated and correlated states are equated as the same.

In the Gutzwiller approximation the postulated wave function has the form
\begin{equation}
|\Psi\rangle=\prod_{i}\left[1-(1-g)n_{i\uparrow}n_{i\downarrow}\right]|\Psi\rangle\equiv
P_{G}|\Psi_{0}\rangle,
\label{11}
\end{equation}
where $|\Psi_{0}\rangle$ represents the Fermi sea of uncorrelated (usually noninteracting) electrons, and $g$ is a variational parameter equal to zero when the double occupancies are excluded. By applying this type of wave function to the narrow-band Hamiltonian, we can obtain the ground state energy $E_{G}/N$ per site of the spin polarized state in the form
\begin{equation}
\frac{E_{G}}{N}=\sum_{\sigma}q_{\sigma}\left(d,n_{\sigma}\right)\bar{\epsilon}_{\sigma}+Ud^{2},
\label{12}
\end{equation}
where $d^{2}\equiv \langle n_{i\uparrow}n_{i\downarrow}\rangle$, $\bar{\epsilon}_{\sigma}$ is the average band energy of particles with spin $\sigma$ (per site), $n_{\sigma}$ is the corresponding average number of particles with spin $\sigma$, and
\begin{equation}
q_{\sigma}=\frac{\left\{\left[\left(n_{\sigma}-d^{2}\right)\left(1-n+d^{2}\right)\right]^{1/2}
+ \left[d^{2}\left(n_{\bar{\sigma}}-d^{2}\right)\right]^{1/2}\right\}^{2}}{n_{\sigma} \left(1-n_{\sigma}\right)}
\label{13}
\end{equation}
is the so-called band narrowing factor which represents a renormalization factor, $0\leqslant q_{\sigma}\leqslant 1$, of bare band energy under the influence of correlations (in the Hartree-Fock limit $d^{2}=\langle n_{i\sigma}\rangle \langle n_{i\bar{\sigma}}\rangle$ or equivalently, $q_{\sigma}=1$). Alternatively, one can write down (\ref{12})as an expectation value for the bare wave function $|\Psi_{0}\rangle$ of the following effective single-particle Hamiltonian \cite{a11}:
\begin{equation}
\mathcal{H}_{GA}=\sum_{ij\sigma}q_{\sigma}(n,d,m) t_{ij}c_{i\sigma}^{\dagger}c_{j\sigma}+NUd^{2},
\label{14}
\end{equation}
where $m=\sum_{\sigma}\sigma n_{\sigma}$ is the spin magnetic moment per site. This effective Hamiltonian contains two terms: the renormalized hopping and the expectation value of the Hubbard term. The replacement of the Hubbard term by its expectation value means that we consider the single-particle propagation in a frozen configuration of the double occupied sites, which at the end is optimized by minimizing $E_{G}$ with respect to $d$.

Hamiltonian (\ref{14}) can thus be easily diagonalized and the corresponding free-energy functional $\mathcal{F}$ constructed in the standard manner has the form:
\begin{equation}
\mathcal{F}^{(GA)}=-k_{B}T\sum_{\pmb{k}\sigma}\ln\left[1+e^{-\beta\left(E_{\pmb{k}\sigma} -\mu\right)} \right]+NUd^{2}+\mu N.
\label{15}
\end{equation}
The quantities $m$ and $\mu$ can be calculated either by minimizing this functional for selected $n, T$, and $U/W$ or alternatively, by writing down selfconsistent equations for them, i.e.
\begin{equation}
n=\frac{1}{N}\sum_{\pmb{k}\sigma}\langle a_{\pmb{k}\sigma}^{\dagger}a_{\pmb{k}\sigma}\rangle;\;\;\;
m=\frac{1}{N}\sum_{\pmb{k}\sigma}\sigma\langle a_{\pmb{k}\sigma}^{\dagger}a_{\pmb{k}\sigma}\rangle.
\label{16}
\end{equation}
Obviously, $\langle a_{\pmb{k}\sigma}^{\dagger}a_{\pmb{k}\sigma}\rangle\equiv \bar{n}_{\pmb{k}\sigma}$ is the average occupancy - the Fermi function.

In the executing of either of the two above procedures one can show explicitly that they do not yield the same results. This means that the fundamental principle of Bogoliubov, which holds for the Hartree-Fock approximation, is not obeyed here. This problem can be traced to the presence of renormalization factor
$q_{\sigma}$ in either $E_{G}$ or $\mathcal{F}$, since then $\sum_{\pmb{k}\sigma} \frac{\partial q_{\sigma}(n,m,d)}{\partial m}\epsilon_{\pmb{k}} \langle n_{\pmb{k}\sigma}\rangle\neq 0.$ This very important difficulty has been omitted in almost all papers utilizing the Gutzwiller approximation.

To overcome this difficulty, we have proposed \cite{a5} that the renormalized Hamiltonian (\ref{14}) has to be supplemented  with additional constraints, expressed with the help of Lagrange-multiplier method when calculating the averages of the type (\ref{16}), providing an appropriate condition minimum of thus corrected Landau functional $\mathcal{F}$. Note that such modification will concern any renormalized in this manner mean-field Hamiltonian (e.g. the above \emph{t-J} or Anderson-Kondo models) as it contains nonanalytic renormalization factor $q_{\sigma}$. In this manner, \emph{the statistical consistency} of the whole approach (in this sense of a correct statistical physics) is guaranteed, i.e., the self-consistent equations and the variational minimization provide the same results. Explicitly, in the situation when we consider normal state, i.e., when only $m$ and $\mu$ (or $m$ and $n$) appear as a thermodynamic variables, we define now the effective Hamiltonian as fo
 llows
\begin{eqnarray}
\widetilde{\mathcal{H}}\equiv \mathcal{H}_{GA}-\lambda_{m}\left(\sum_{\pmb{k}\sigma}\sigma n_{\pmb{k}\sigma}-\sum_{\pmb{k}\sigma}\sigma \langle n_{\pmb{k}\sigma}\rangle \right) \nonumber \\
-\lambda_{n}\left(\sum_{\pmb{k}\sigma}n_{\pmb{k}\sigma}-\sum_{\pmb{k}\sigma}\langle n_{\pmb{k}\sigma}\rangle \right),
\label{18}
\end{eqnarray}
where the Lagrange multipliers $\lambda_{m}$ and $\lambda_{n}$ play the role of extra (global) molecular fields to be calculated also from the corresponding minimum condition for $\mathcal{F}$, what guarantees automatically the statistical consistency as one can see by imposing $\partial\mathcal{F}/\partial\lambda_{m}= \partial\mathcal{F}/\partial\lambda_{m}=0$.

One should note that the variational parameter $d$ \textbf{is not} a thermodynamic variable, so the corresponding constraint does not appear. Nonetheless, in such modified formulation all the variables appearing in (\ref{15}) are calculated variationally and thus the approach becomes self-consistent and self-contained. For the sake of simplicity we do not elaborate in detail of the approach using the Fukushima approach \cite{a13, a14, a15}. However, we discuss some of the results obtained by using this particular method.

\subsection{Statistical consistency for \emph{t-J} model and Fukushima variational wave function (SCA)}

The statistical consistency conditions have been implemented practically simultaneously to the two models interesting us in the context of unconventional superconductivity: the extended \emph{t-J} model with real space pairing induced by the antiferromagnetic kinetic exchange \cite{a5,a13,a14} as well as to the Anderson-Kondo model with hybrid real space pairing induced by the Kondo coupling \cite{a15,a19}. Both of these types of pairing have been introduced in the preceding Section.

The situation in the \emph{t-J} model has been additionally modified by taking into account a modified Gutzwiller-type wave function introduced by Fukushima \cite{a10}. In the latter approach, the Gutzwiller wave function (\ref{11}) in the strong-correlation limit ($g=0$) is replaced by
\begin{equation}
|\Psi_{F}\rangle=\prod_{i}\left(\lambda_{i\uparrow}\right)^{n_{i\uparrow}/2} \left(\lambda_{i\downarrow} \right)^{n_{i\downarrow}/2}\left(1-n_{i\uparrow}n_{i\downarrow}\right)|\Psi_{0}\rangle\equiv
P_{F}|\Psi_{0}\rangle.
\label{19}
\end{equation}
Such modification of the Gutzwiller projector $P_{G}$ allows us imposing the condition, that uncorrelated average number of electrons per site (on each site "i") and with spin is equal the corresponding actual average computed within the scheme. In other words,
\begin{equation}
\langle n_{i\sigma}\rangle_{F}=\langle n_{i\sigma}\rangle,
\label{20}
\end{equation}
where subscript "$F\,$" mean evaluation with the wave function (\ref{9}). This conditions represent additional self-consistency requirement. In effect, as well will see, some of the renormalization factors are the same, some are different.

Two important practical points concerning this method should be emphasized. First, since the ground state energy  is expressed as
\begin{equation}
E_{G}\equiv\frac{\langle\Psi_{F}|\widetilde{\mathcal{H}}|\Psi_{F}\rangle}{\langle\Psi_{F}| \Psi_{F}\rangle},
\label{21}
\end{equation}
and $| \Psi_{0}\rangle$ represents an uncorrelated state, the multiple-fermion-operator expectation values contained in (\ref{20}) can be evaluated by factorizing them into those containing only pairs of operators, each evaluated for uncorrelated state (a Wick-type contraction!). This is somewhat cumbersome procedure, so we present only the final result:
\begin{equation}
\lambda_{i\sigma}=\frac{1-\langle n_{i\sigma}\rangle}{1-\langle n_{i}\rangle},
\label{22}
\end{equation}
as well as the form of the effective \emph{t-J} Hamiltonian:
\begin{eqnarray}
\stackrel{\approx}{\mathcal{H}}&=&
\frac{\langle\Psi_{0}|P_{F}\mathcal{\widetilde{H}}P_{F}|\Psi_{0}\rangle} {\langle\Psi_{0}|P_{F}^{2}|\Psi_{0}\rangle}\nonumber\\
&-&\sum_{i}\lambda_{i}^{(n)}\left(\sum_{\sigma} a_{i\sigma}^{\dagger}a_{i\sigma}-\left\langle a_{i\sigma}^{\dagger}a_{i\sigma}\right\rangle\right) \nonumber\\
&-&\sum_{\langle ij\rangle\sigma} \lambda_{ij}^{(\chi)}\left[\left(c_{i\sigma}^{\dagger}c_{j\sigma} -\chi_{ij}\right)+H.c.\right] \nonumber\\
&-&\sum_{\langle ij\rangle\sigma} \lambda_{ij}^{(\Delta)}\left[\left(c_{i\bar{\sigma}}c_{j\sigma} -\Delta_{ij}\right)+H.c.\right],
\label{23}
\end{eqnarray}
where the effective \emph{t-J} Hamiltonian SCA $\mathcal{\widetilde{H}}$ is expressed by a corresponding expression SCA, which will not be reproduced in detail here (see \cite{a13}).  However, on should note that in (\ref{23}) we take in this method an expectation value of $\langle\mathcal{\widetilde{H}}\rangle$. Thus the operator part of the Hamiltonian $\stackrel{\approx}{\mathcal{H}}$ is composed of the constraints. Such trick is analogous to that leading to the effective single-particle Hamiltonian (\ref{16}), where the expectation value of the interaction part is taken. Note also, that we evaluate $\langle P_{F}\mathcal{\widetilde{H}}P_{F}\rangle$; as well as $\langle P_{F}^{2}\rangle$ as expectation for the noninteracting state $|\Psi_{0}\rangle$. So, we can utilize the  Wick-type factorization of the operators in the real space representation.

\subsection{Statistically consistent approximation for the Anderson-Kondo lattice model}

Very recently, we have extended \cite{a19} the Fukushima approach to the periodic Anderson-Kondo Hamiltonian (\ref{10}). First, we have extended our former results \cite{a20} for heavy-fermion systems to the case with nonzero applied magnetic field. Second, the work is progressing on the hybrid real-space pairing and its coexistence with magnetism. The latter work \cite{a21} extends our earlier work on unconventional superconductivity with hybrid pairing in a lower hybridized band and in the Gutzwiller approximation. This work will be elaborated elsewhere \cite{a15,a19}. Also, it would be desirable to analyze high-temperature superconductivity within $d-p$ hybridization included explicitly and within the SCA approach.

\subsection{Statistically consistent approximation: A brief outlook}

It is very important to note that the constraints introduced and composing an essence of the SCA scheme are the same, at least for the normal state, with the more involved auxiliary (slave)-boson type of approach in the saddle-point approximation \cite{a5}. However, our approach contains no auxiliary ("ghost") Bose condensed fields, which introduce spurious phase transitions. Second, the present approach is quite natural on physical grounds, as well as extends the long tradition of introducing a molecular field as a conjugate variable to each introduced order parameter, here in the situation with a complex ordering. Also, as one can see already from the complex form of the mean-field Hamiltonian, that it represents a quantum liquid state intermediate between \emph{the Landau Fermi-liquid state} and what is called a \emph{non-Fermi} liquid state, in addition to providing the Mott or selective-orbital-Mott localized states, as quite unconventional, as we discuss it elsewhere.

In summary, the SCA incorporates the slave-boson approach into a statistically-consistent single-particle mean-field theory with a complex formal structure involving a number of order parameters and associated with them mean fields. Once the form of a variational wave function is selected, all the remaining analysis can be systematic in the sense of many-particle perturbation theory \cite{a22}.

\section{Physical properties: superconducting and magnetic states}

The most important feature of any theory is to provide a quantitative, or at least a coherent semiquantitative, description of the relevant electron states and properties and in particular, a phase diagram involving physically plausible phases that given fermionic model allows for, at least in the in mean-field approximation. Below we characterize separately recent results for the \emph{t-J} model and for the hybridized (two-orbital) Anderson-Kondo model, in both cases involving the stable phases.

\subsection{High-temperature, single plane superconductivity within \emph{t-J} model in SCA approximation}

To determine the role of different terms we rewrite the Hamiltonian (\ref{2}) in the following form
\begin{equation}
\widetilde{\mathcal{H}}_{t-J}\equiv\sum_{ij\sigma}\!^{'}t_{ij}b_{i\sigma}^{\dagger}b_{j\sigma}+
\sum_{\langle ij\rangle}J_{ij}\pmb{S}_{i}\cdot\pmb{S}_{j}-\frac{c_{1}}{4}\sum_{\langle ij\rangle}\widetilde{K}_{ij}\nu_{i}\nu_{j}+c_{2}\mathcal{H}_{3},
\label{24}
\end{equation}
where $c_{1}=c_{2}=1$ and $\widetilde{K}\equiv J_{ij}-4K_{ij}$, $J_{ij}=4t_{ij}^{2}\left(U-K_{ij}\right)$, $\mathcal{H}_{3}$ represents the three-site terms, and $\langle ij\rangle$ means the summation over nearest neighboring pairs, each taken once. The factor $c_{1}$ and $c_{2}$ have been introduced to differentiate between the various contributions to the paired state energy. Also, we consider here only the case with $K_{ij}\equiv0$, i.e. neglect the \emph{intersite} part of the repulsive Coulomb interaction.  The rationale behind the last assumption is that, since we consider a correlated-pair motion as the source of pairing, the repulsive contributions is only one of the contribution, not the decisive one. Additionally, we discuss here only the final results, as the formal analysis has been elaborated in detail elsewhere \cite{a13,a14,a15}. Let us mention only that the result can be represented, by the Hamiltonian (\ref{24}) reduced to the renormalized form
\begin{equation}
\mathcal{H}_{t-J}^{(ren)}=\sum_{ij\sigma}\!^{'}g_{ij}^{t}a_{i\sigma}^{\dagger}a_{j\sigma}+ \sum_{ij}\!^{'}g_{ij}^{\Delta}J_{ij}\Delta_{ij}^{*}\Delta_{ij}+E_{0}+
\ldots
\label{25}
\end{equation}
where $g_{ij}^{t}$ and $g_{ij}^{\Delta}$ are the corresponding factors. The form of $g_{ij}^{t}$ follows from a detailed procedure which leads to the renormalized hopping in the correlated ($|\Psi_{F}\rangle$) state:
\begin{eqnarray}
\langle a_{i\sigma}^{\dagger}a_{j\sigma}\rangle_{C}&\simeq&\sqrt{\frac{1-\bar{n}_{i}}{1-\bar{n}_{i\sigma}}}
\;\sqrt{\frac{1-\bar{n}_{j}}{1-\bar{n}_{j\sigma}}} \nonumber \\
&\cdot&\left[\chi_{ij\sigma}-\chi_{ij\bar{\sigma}} \frac{\chi_{ij\sigma}\chi_{ij\bar{\sigma}}^{*}+\Delta_{ij}\Delta_{ij}^{*}} {\left(1-\bar{n}_{i\bar{\sigma}}\right) \left(1-\bar{n}_{j\bar{\sigma}}\right)}
\right],
\label{26}
\end{eqnarray}
where $\chi_{ij\sigma}\equiv\langle a_{i\sigma}^{\dagger}a_{j\sigma}\rangle$ is the hopping amplitude in the uncorrelated state and therefore for the paramagnetic state with real gap parameter $\Delta_{ij}\equiv\langle c_{i\bar{\sigma}}^{\dagger}c_{j\sigma}\rangle=\Delta_{ij}^{*}$ we have that
\begin{eqnarray}
g_{ij}^{t}=\sqrt{\frac{1-\frac{\bar{n}_{i}}{2}}{1-\bar{n}_{i}}}\;
\sqrt{\frac{1-\frac{\bar{n}_{j}}{2}}{1-\bar{n}_{j}}}\;
\left[
1-\frac{\chi_{ij}^{2}+\Delta_{ij}^{2}}{\left(1-\frac{\bar{n}_{i}}{2}\right) \left(1-\frac{\bar{n}_{j}}{2}\right)}
\right],
\label{27}
\end{eqnarray}
and
\begin{eqnarray}
g_{ij}^{\Delta}=\sqrt{\frac{1-\frac{\bar{n}_{i}}{2}}{1-\bar{n}_{i}}}\;
\sqrt{\frac{1-\frac{\bar{n}_{j}}{2}}{1-\bar{n}_{j}}}\;
\left[
1+\frac{\chi_{ij}^{2}+\Delta_{ij}^{2}}{\left(1\frac{\bar{n}_{i}}{2}\right) \left(1-\frac{\bar{n}_{j}}{2}\right)}
\right].
\label{28}
\end{eqnarray}
Here $\bar{n}_{i}\equiv\langle n_{i}\rangle$.
These renormalization factors reduce to the standard Gutzwiller-ansatz form for spatially homogeneous case, $\bar{n}_{i}=\bar{n}_{j}=n$. Here, to first approximation $g_{ij}^{t}\simeq g_{ij}^{\Delta}$, i.e. the formation of the pairing requires electron itineracy. Note, that strictly speaking
\begin{equation}
\left\langle\Psi_{F}\left|\mathcal{\widetilde{H}}_{t-J}\right|\Psi_{F}\right\rangle= \left\langle\Psi_{0}\left|\widetilde{\mathcal{H}}_{t-J}^{(ren)}\right|\Psi_{0}\right\rangle.
\label{29}
\end{equation}
Therefore, the exchange and other interaction parts not written down explicitly, should be decoupled in the mean field manner using all possible (assumed as nonzero) contracted pair-operator averages. In effect, we can diagonalize thus approximated Hamiltonian (\ref{24}) with the decoupled terms and supplemented with the constraints. The resultant renormalized hopping amplitude $T_{\tau}$ and the gap parameters $D_{\tau}$ in the $\tau$-th direction of the lattice, have the following forms:
\begin{eqnarray}
 T_{\tau}&\equiv & t_{\tau}g_{\tau}^{t}+\frac{3}{4}J_{\tau}g_{\tau}\chi_{\tau}+\lambda_{\tau}^{(\chi)},
\label{30}\\
D_{\tau}&\equiv & \frac{3}{4}J_{\tau}g_{\tau}^{J}\Delta_{\tau}+\lambda_{\tau}^{(\Delta)},
\label{31}
\end{eqnarray}
where we have included only the terms written down explicitly in (\ref{25}). Also, we have assumed that $\chi_{x}=\chi_{y}=\chi$ and $\Delta_{x}=-\Delta_{y}=\Delta$,
the second relation being for the superconducting gap of the $d$-wave form.
\begin{figure}
\includegraphics[width=0.45\textwidth]{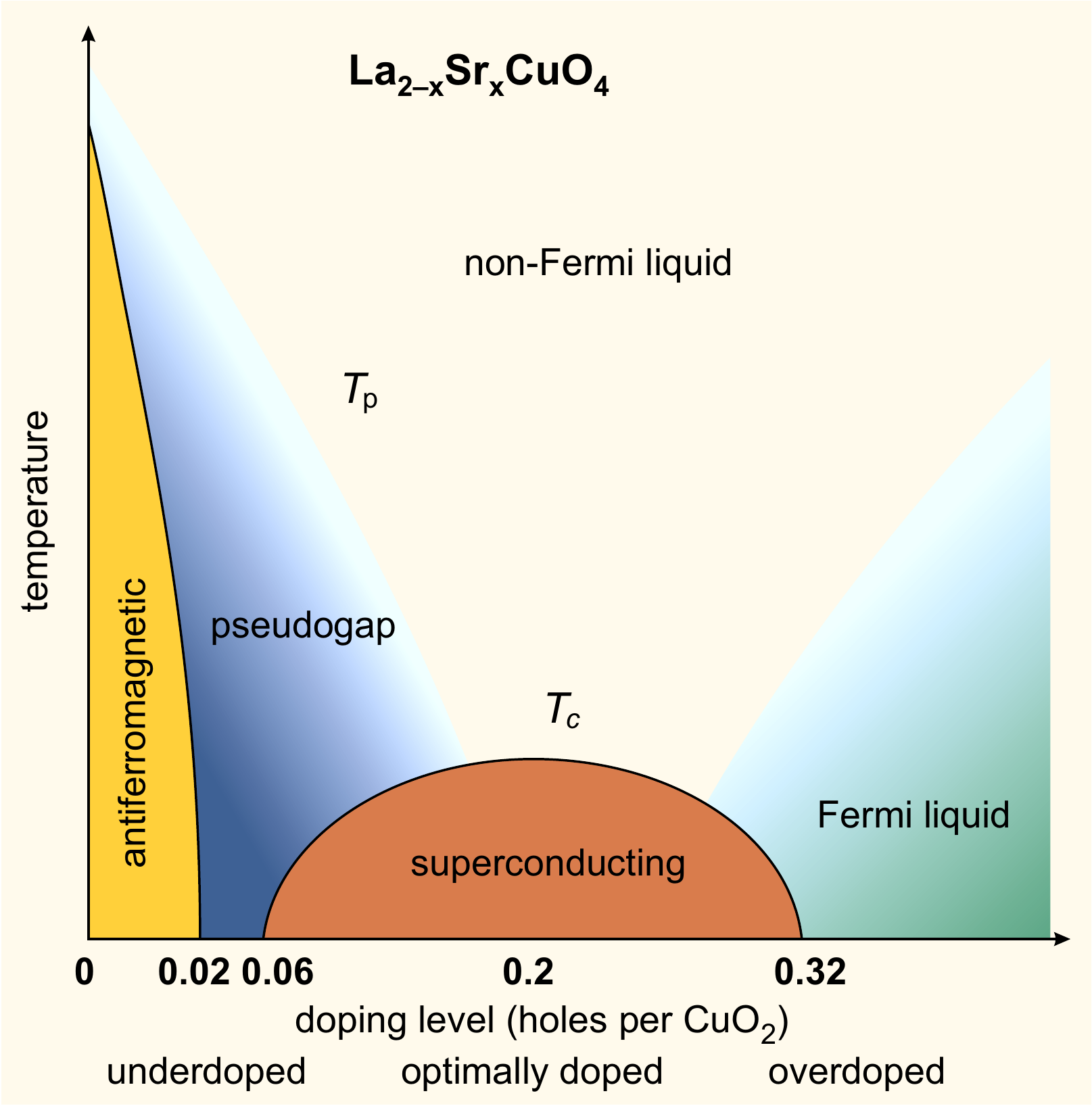}
\vspace{1cm}{
\includegraphics[width=0.45\textwidth]{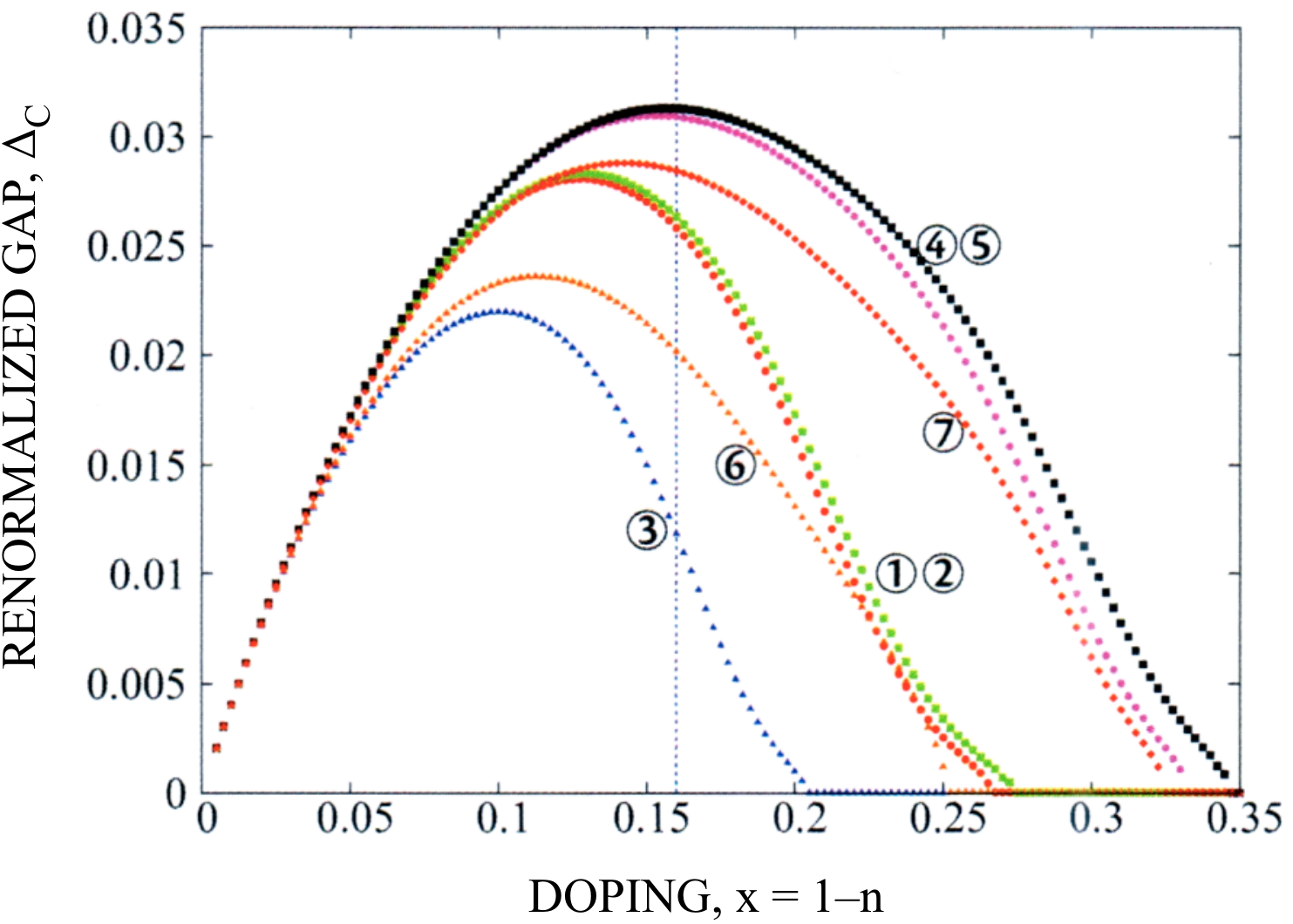}}
\caption{(Color online) Top: overall phase diagram for $La_{2-x}Sr_{x}Cuo_{4}$ on the temperature $T$ - doping $x$ plane. Bottom: Doping dependence of the renormalized superconducting order parameter $\Delta_{c}$. The curves 1-7 are explained in detail in Ref. [11] and correspond to various methods of solving Hamiltonian (2). The most important feature of the solution is the appearance of the upper critical concentration for the $d$-wave superconductivity disappearance. The vertical line defines roughly the optimal doping. }
\end{figure}

In Fig. 3a we plot the overall phase diagram for the high temperature superconductor $La_{2-x}Sr_{x}Cu_{4}$. The corresponding theoretical phase diagram containing the superconducting part of the phase diagram is exhibited in Fig. 3b. Few features of theoretical phase diagram should be noted. First, the superconducting disappears only in the Mott insulating limit $n\rightarrow 1$ (doping $x\equiv 1-n\rightarrow 0$). This inadequacy of our approach are caused by the circumstance, that we have not included in our analysis either the onset of antiferromagnetism or the atomic disorder induced by the $Sr$ doping. Second problem with our approach is connected with the absence of \emph{the pseudogap appearance}. This is because we neglect here the phase fluctuations of the order parameter $\Delta_{ij}$. Nonetheless, the approach contains the two very attractive features. First, we show clearly, that there exist an upper critical concentration for the disappearance of superconductivity. The various curves, which were calculated with different approximations in the Hamiltonian (as discussed in detail in \cite{a13}), show clearly that the upper critical concentration is the doping regime $x=0.22-0.35$, in agreement with the experimental data for various single-plane cuprates. One should state right away, that the existence of this critical concentration speaks out decisively in the favor of real space pairing, as with increase of doping the hopping increases, which in conjunction with dilution electrons with $x$ destroys the correlated pair motion. Second, here we have a well defined Fermi-liquid state, particularly in the so-called \emph{overdoped regime} (to the right of the vertical line in Fig. 3b). The real space involves a correlated-pair motion and has been animated by us recently \cite{a23}. To put it bluntly, the pairing discussed here does not involve any intermediate boson such as paramagnon. Finally, the magnitude of the renormalized gap $\Delta_{c}\equiv g^{\Delta}\langle\Delta_{ij}\rangle$ is in the unit of the first hopping integral $|t_{1}|$  which is taken in the range $0.35-0.40 eV$ ($|t_{1}|/J=3$ is assumed). Therefore, the maximal value of $\Delta\sim 100-120K$ is obtained for the uppermost curve, a quite reasonable value in view of the mean-field nature of our estimate. An elementary interpretation  of the superconducting "dome" shape depicted in Fig. 3 is based on the circumstance that here renormalized hopping amplitude $\sim |t|x$ is comparable to $J$ \cite{a23}.

\begin{figure}[h!]
\begin{center}  
\rotatebox{270}{\scalebox{0.32}{\includegraphics{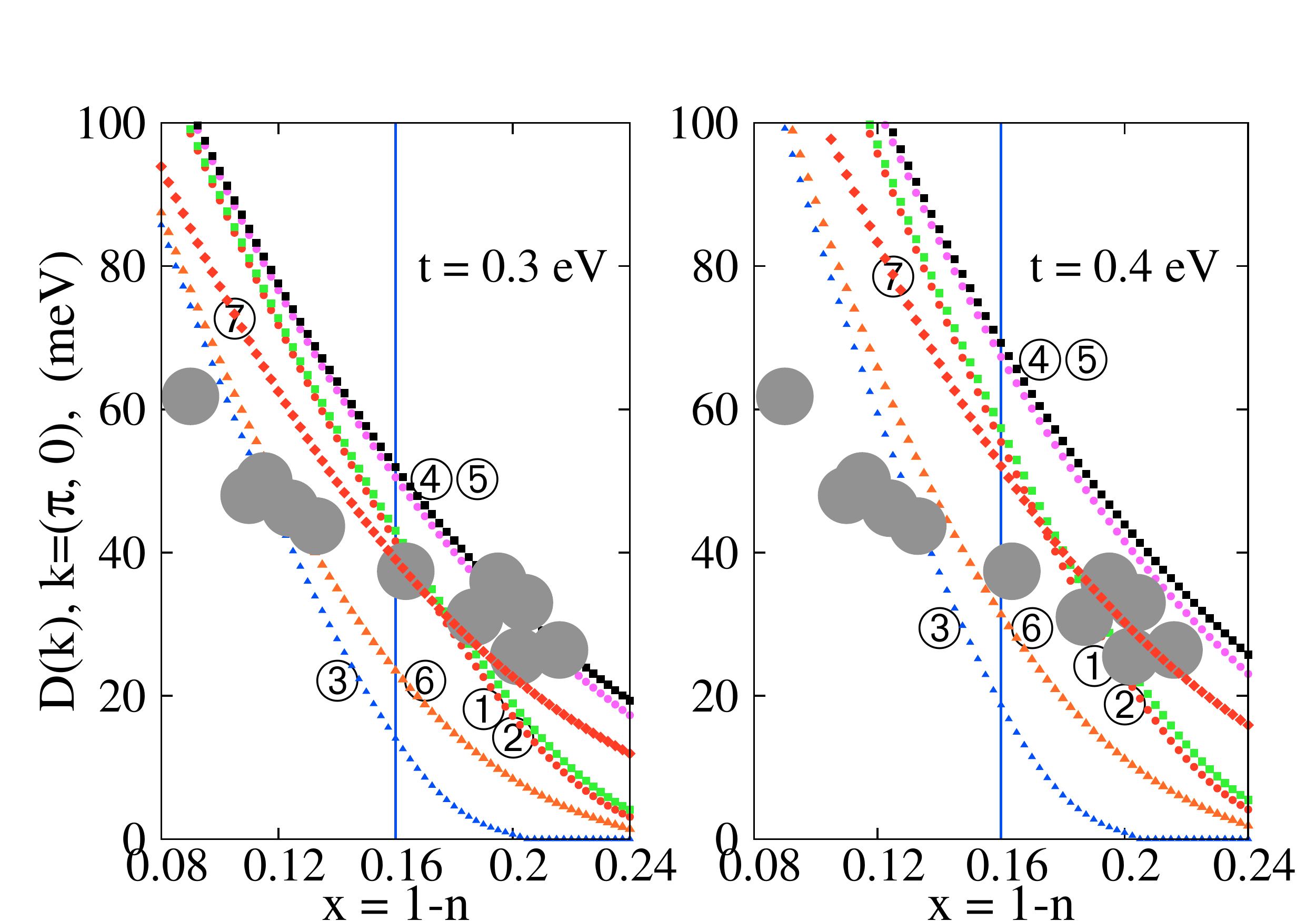}}}
\end{center}
\caption{(Color online) Doping dependencies of the  SC gap $D_{\pmb{k}}$ at $\pmb{k} = (\pi, 0)$ for  cases 1-6 and for $t^{\prime}/t = -0.27$, and $J/|t| =0.3$ (filled diamonds). Large filled circles - experimental data. For a detail discussion see Ref. \cite{a13}.}
\end{figure}
\begin{figure}[h!]
\begin{center} 					
\rotatebox{270}{\scalebox{0.32}{\includegraphics{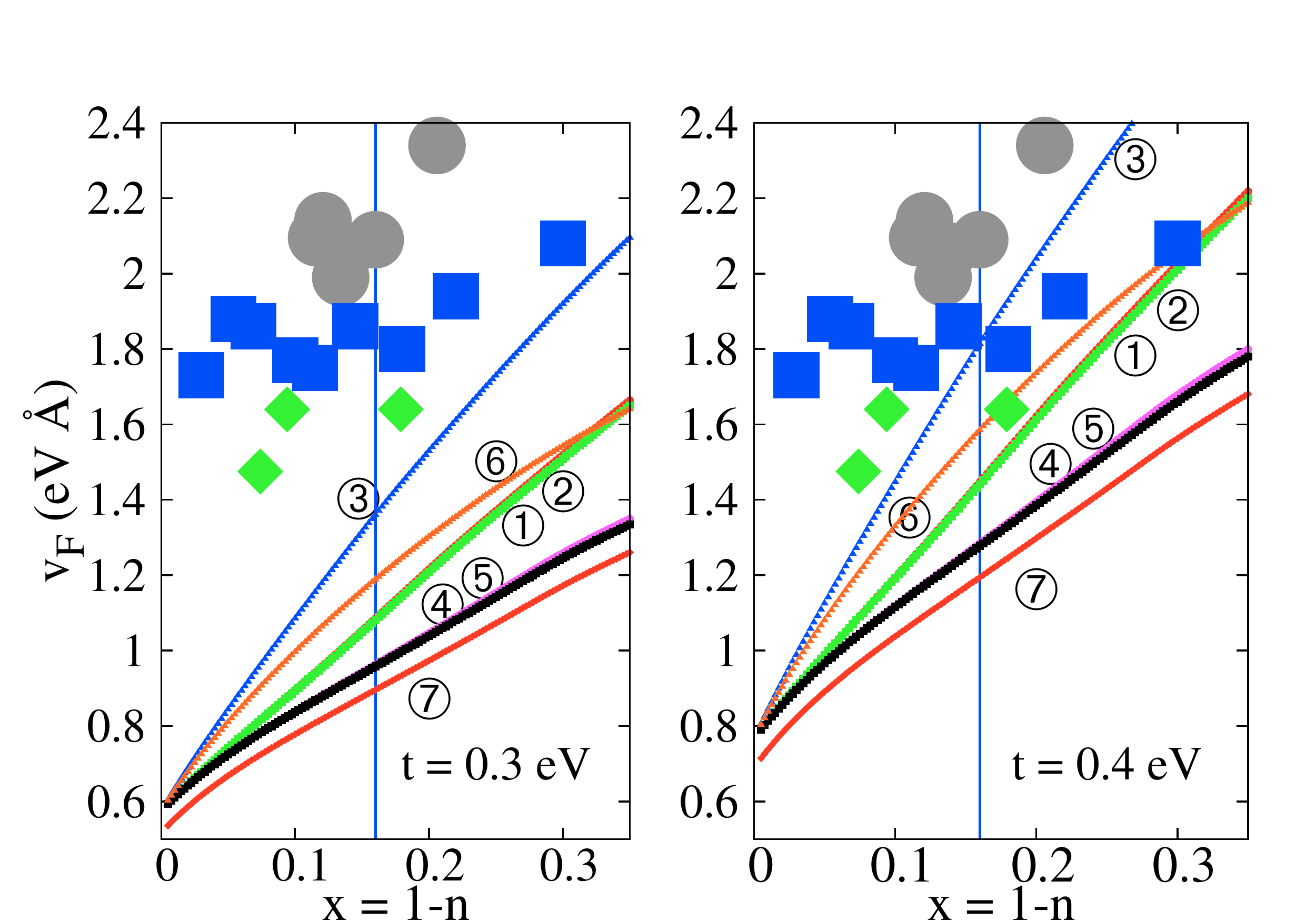}}}
\end{center}
\caption{(Color online) Doping dependence of Fermi velocity in the nodal ($  (0, 0) \to (\pi, \pi)$) direction. For a detail discussion see Ref. \cite{a13}.}
\end{figure}

In Figs. 4 and 5 we display respectively, the doping dependence of the quasiparticle energy in the antinodal direction ($\pmb{k}=[\pi,0]$) and the Fermi velocity in the nodal $([0,0]\rightarrow [\pi,\pi])$ direction. The points represent various experimental results for the single-planar systems \cite{a13}. Our analysis represents a more precise version of that presented in Ref. \cite{a14}.

From Figs. 3-5 we can see that our results express in a semiquantitative manner the trend of the corresponding experimental data for overdoped systems (to the right of the vertical line). However, it must be reiterated again, that the phase diagram does not contain the crossover (pseudogap) line, cf. Fig. 3b, as well as provide a wrong tendency of the Fermi velocity $v_{F}$, i.e., its almost constant value in the underdoped regime. This most probably means that pocket of Fermi-surface (the arcs \cite{a25}) near the nodal direction survive almost untouched in underdoped regime. The situation in the underdoped regime requires a generalization of the present SCA approach (some hints to that generalization are provided in Appendix E).

One more interesting feature of our results should be noted. Namely, the temperature dependence of the superconducting gap magnitude $\Delta_{C}\equiv\Delta_{C}(T)$ follows to a very good approximation the BCS dependence in the version proposed by Rickayzen \cite{a26}, which in our context
\begin{equation}
\frac{\Delta_{C}(T)}{\Delta_{C}(0)}=\tanh\left(\frac{\Delta(T)}{\Delta(0)}\;
\frac{T_{C}}{T}\right),
\label{32}
\end{equation}
with relative temperature $t\equiv T/T_{C}$. The comparison our formula (\ref{32}) is illustrated numerically in Fig.6. This is an amazing result given the complicated nature of the self-consistent equations leading to the numerical results. For the sake of completeness we derive the above equation in Appendix B. We also show there that $2\Delta_{C}(0)/k_{B}T_{C}=4$.
\begin{figure}[h!]
\begin{center} 					
\rotatebox{270}{\includegraphics[width=0.3\textwidth]{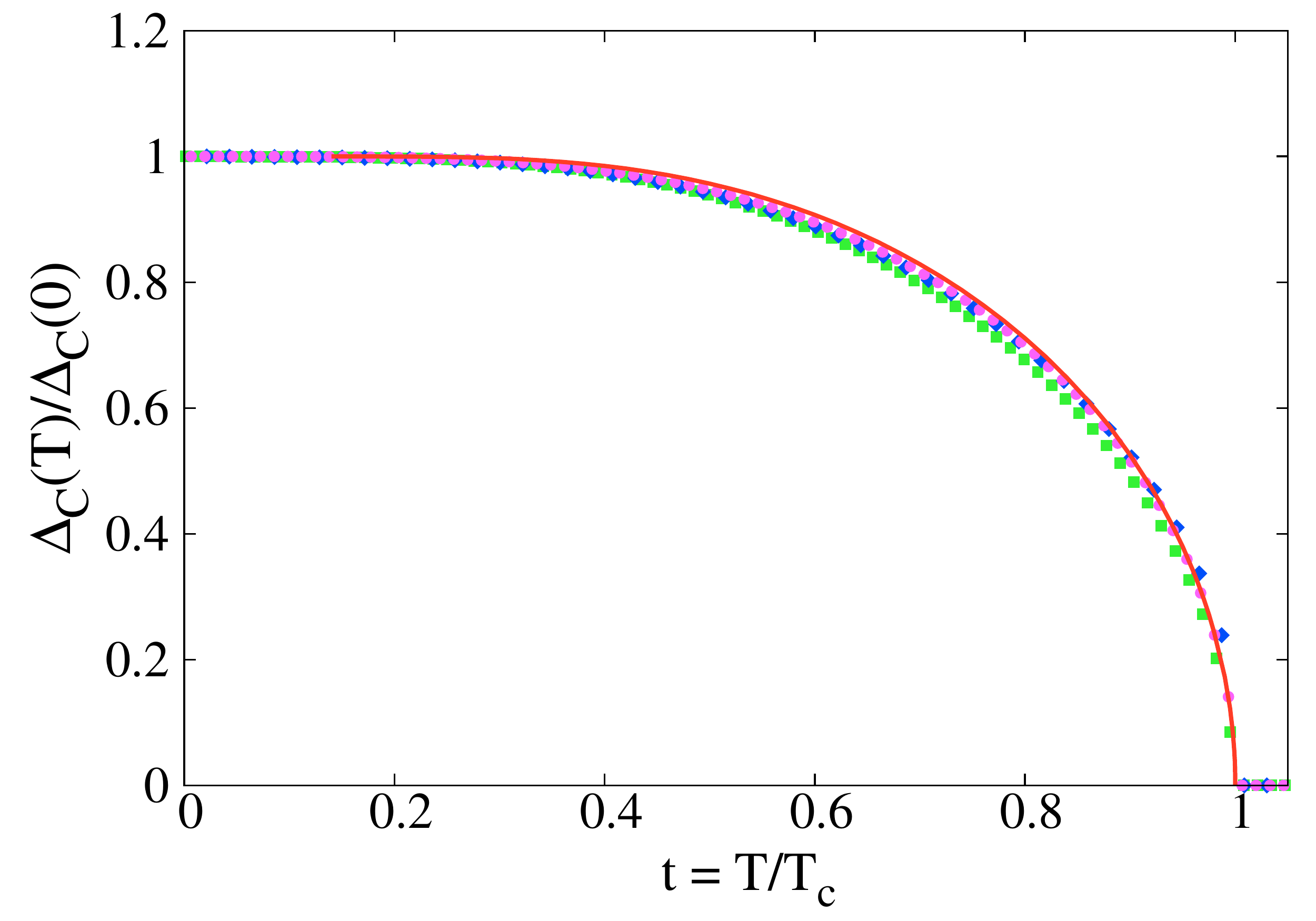}}
\end{center}
\caption{(Color online) Reduced temperature dependence of renormalized superconducting order parameter  $\langle \hat{\Delta} \rangle_C \equiv \Delta_C(T)$ for various doping levels: squares: $x=0.125$, solid circles: $x=0.25$, diamonds: $x=0.3$ are all for $c_1 =1$, $c_2 =0$, and $t^{\prime}/t = -0.25$. Solid line: BCS result (\ref{32}).}
\end{figure}

\subsection{\emph{t-J} model: coexistence of magnetism and superconductivity}

A complete analysis of the \emph{t-J} model requires determination of the Fermi-surface evolution with doping within this almost localized Fermi-liquid picture, along the lines presented in \cite{a13}. Furthermore, one should incorporate the magnetic phases into the phase diagram presented in Fig. 3b. The last type of analysis has been performed very recently \cite{a18,a14} and the results are presented in Fig. 7, where the same type of approach as above has been applied with one addition feature.
Namely, a mixture of the spin-singlet and the so-called staggered spin triplet states appears when the antiferromagnetism coexists with superconductivity \cite{a27}. The appearance of the triplet component of the gap is imminent in that situation even though we have only the singlet-pairing interaction or equivalently, when only the antiferromagnetic exchange interaction in present. This is because of the following reason, illustrated in Fig. 7. Namely, if we have two-sublattice antiferromagnet, the pairing amplitude $\Delta_{A}\equiv\langle c_{i\uparrow} c_{j\downarrow}\rangle$ between the spin majority electrons sublattices ($i \in A, j \in B $) should be different (larger) than that on the corresponding quantity $\Delta_{B}\equiv\langle c_{i\downarrow}c_{j\uparrow}\rangle$ for the minority-spin electrons. Additionally, we would like to describe the $d$-wave superconductivity, expressed in the real-space language, which  amounts to postulating the following form of the gaps
\begin{figure}
\begin{center}
\includegraphics[width=0.35\textwidth]{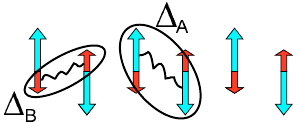}
\end{center}
\caption{(Color online) Spin-majority (blue, bigger arrows) and spin-minority (red, smaller arrows) electron spins in a system with the AF order and real-space superconducting gaps. $\Delta_{A}$ binds two spin-majority electrons, and $\Delta_{B}$ binds two spin-minority electrons and therefore, there is a priori no reason for these two gaps to coincide (as would be the case for no staggered $\pi$-triplet component). In other words, the two distinct gaps make effectively the $\ua-\da$ and $\da-\ua$ pairing components of the opposite-spin pairs distinguishable.}
\end{figure}
\begin{eqnarray}
\Delta_{ij}\equiv \langle c_{j\downarrow} c_{i\uparrow} \rangle=\left\{
{{\tau_{ij}\Delta_{A} \;\;\;\mbox{for} \;\;\; i\in A}
\atop {\tau_{ij}\Delta_{B}\;\;\; \mbox{for}\;\;\; i\in B}}
\right.,
\end{eqnarray}
where $\tau_{ij}=+1$ for the n.n. pair $\langle ij\rangle$ along $x$ and $\tau_{ij}=-1$ along $y$ axis, respectively. One can note immediately, that if $\langle c_{j\downarrow} c_{i\uparrow}\rangle$ and $\langle c_{j\uparrow} c_{i\downarrow}\rangle$
averages differ in magnitudes, then we can have a mixture of singlet and triplet pairing, as in e.g. singlet-pairing case they would be of opposite sign and for pure triplet the sign would be the same. In effect, the unrenormalized gap parameter $\Delta_{ij}$ can in the present situation be decomposed into the singlet ($\Delta_{ij}^{(S)}$) and the triplet ($\Delta_{ij}^{(T)}$) parts according to a prescription
\begin{equation}
\Delta_{ij}=\Delta_{ij}^{(S)}+\Delta_{ij}^{(T)}e^{i\pmb{Q}\cdot\pmb{R}_{i}},
\end{equation}
with
\begin{equation}
\Delta_{ij}^{(S)}\equiv\frac{1}{2}\left(\Delta_{A}+\Delta_{B}\right),\;\; \Delta_{ij}^{(T)}\equiv\frac{1}{2}\left(\Delta_{A}-\Delta_{B}\right)e^{i\pmb{Q}\cdot\pmb{r}_{i}},
\label{35}
\end{equation}
where $\pmb{Q}=(\pi, \pi)$ is the superlattice vector in the case of simple two-sublattice AF ordering in two dimensions. The amazing feature of the present representation of the gap in the form  (\ref{35}) is that the triplet component is present even though there is no explicit spin-triplet-pairing inducing interaction in the Hamiltonian. The detailed analysis has been performed in \cite{a14}, where the corresponding statistical-consistency constraints and the minimization of the (appropriate free-energy) functional has been carried out. One should note that in the present situation we encounter, for the first time in this approach, the spatially modulated occupancy, i.e, it has the form:
\begin{equation}
\left\langle n_{i\sigma}\right\rangle \equiv \left\langle\Psi_{0}|n_{i\sigma}|\Psi_{0}\right\rangle=\frac{1}{2}\left(n+\sigma m_{F}+\sigma m_{AM}e^{i\pmb{Q}\cdot\pmb{R}_{i}}\right),
\end{equation}
where, as before, $n$ is the band filling ($0\leqslant n\leqslant 1$), $m_{FM}$ is a ferromagnetic (homogeneous) spin-moment component, and $m_{AF}$ is the antiferromagnetic (staggered, sublattice) moment component. The results, encompassing a simultaneous optimization of a system of 11 algebraic equations for the case of square-lattice case, is shown in Fig. 8 in the form of the phase diagram on the applied magnetic field $h$ (in units of $|t|$) - band filling plane. The microscopic parameters taken in that computation are listed in that figure.
\begin{figure} 
\begin{center} 					
\rotatebox{270}{\includegraphics[width=0.35\textwidth]{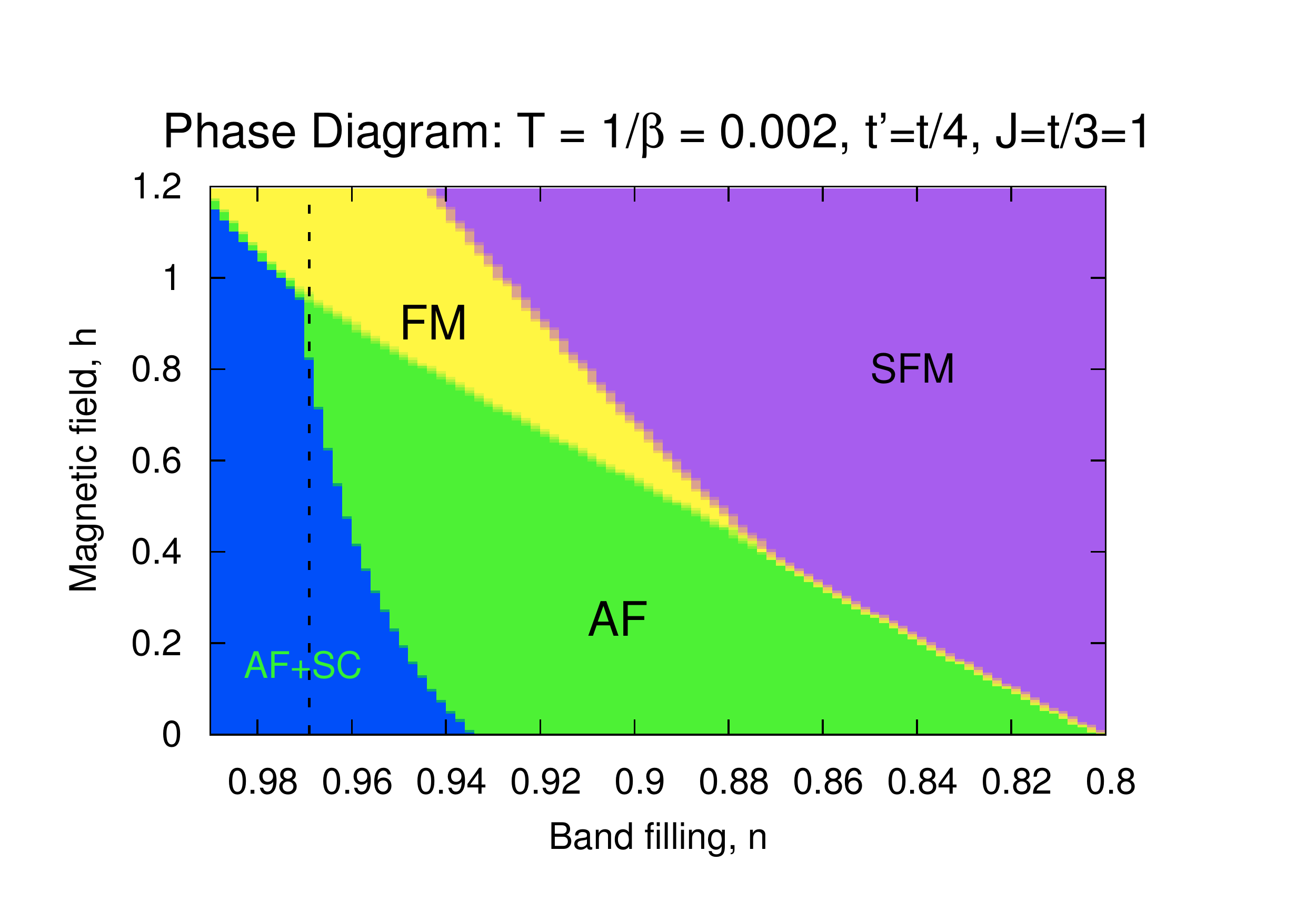}}
\end{center}
\caption{(Color online). Phase diagram on the band filling - magnetic field plane. The phases are labeled as follows: AF+SC - phase with coexisting superconductivity and antiferromagnetism, AF - antiferromagnetic phase, FM - ferromagnetic phase, SFM - saturated ferromagnetic phase (with $m_{FM} = n$). For further analysis we restrict ourselves to $n=0.97$ as marked by the dashed vertical line. No stable pure superconducting solution has been found.}
\end{figure}

Particularly interesting for us is the horizontal line, where we observe the sequence of phases $AF+SC \rightarrow AF \rightarrow PP \rightarrow FM$ with the decreasing band filing. In a way, the results are somewhat disappointing, as we would rather expect, with the increasing $n$, a clear transition from $AF$ Mott insulating state to a pure superconducting state. Here the coexisting phases $AF+SC$ phase are stable up to $x_{c'}=0.0.6$, and the for $x>x_{c}$ the pure two-sublattice AF state becomes stable.

\subsection{Fulde-Ferrell phase in narrow-band limit}

In this Section we overview briefly plausibility of the observations of the Fulde-Ferrell phase in the strongly correlated-electrons systems, as viewed from the point of view of our SGA approach \cite{a28}.

The Fulde-Ferrell-Larkin-Ovchinnikov (FFLO) was proposed theoretically many years ago \cite{a29}. The motivation for our work in this area was the suggestion that this state has been observed in the heavy-fermion system $CeCoIn_{5}$, possibly coexisting with antiferromagnetism \cite{a30}. It was quite a coincident with the experimental observation of spin-dependent heavy electron masses \cite{a31}. On one hand, the appearance of the spin-dependent heavy electron mass was proposed earlier as one of the crucial phenomena for the strong correlated systems \cite{a32}. Therefore, a natural idea appeared to discuss the effect of the spin-dependent masses (SDM) on the FFLO state stability. However, we discuss first briefly the concept of the spin-dependent masses.

\subsection*{Spin-dependent masses}

For a quasiparticle gas of correlated particles, their energy in the applied field $h\equiv\frac{1}{2}g\mu_{B}H_{a}$, when the Landau quantization is neglected, is expressed as \cite{a28}
\begin{equation}
\xi_{\textbf{k}\sigma}=\frac{\hbar^2k^2}{2m\sigma}-\sigma h-\mu-\sigma h_{corr},
\end{equation}
where $h_{corr}$ is the effective field induced by the correlations and spin-dependent mass enhancement $m_{\sigma}/m_{B}$ is
of the form in the limit of $U\rightarrow\infty$ for a single narrow band \cite{a31}
\begin{eqnarray}
\frac{m_{\sigma}}{m_{B}}=\frac{1-n_{\sigma}}{1-n}=\frac{1-n/2}{1-n}-\sigma\frac{\bar{m}}{2(1-n)}
\nonumber \\
\equiv\frac{1}{m_{B}}\left(m_{av}-\frac{\sigma}{2}\Delta m\right),
\end{eqnarray}
where $m_{B}$ is the bare (band) mass and $\bar{m}=n_{\uparrow}-n_{\downarrow}$ is here the system spin polarization. Also, $\Delta m=m_{\downarrow}-m_{\uparrow}$ is the mass difference, while $m_{av}=\left(m_{\uparrow}+m_{\downarrow}\right)/2$ is the average quasiparticle mass, i.e., the mass in the absence of magnetic polarization. It is interesting to note that in the magnetic-saturation limit we recover the band limit for the spin-majority subband, i.e. $m_{\uparrow}/m_{B}=1$, whereas the heavy quasiparticles in the spin-minority band (with $m_{\downarrow}/m_{B}=1/(1-n)$) disappear at the border of magnetic-moment saturation $\bar{m}=n$. Two features are important here: (i)  the masses are high in the almost-localized limit $(1-n)\ll 1$, and (ii) the mass $m_{\downarrow}$ in the spin-minority band is the heaviest, since due to the spin subband occupancy imbalance, these spin-minority quasiparticles scatter very strongly (due to the presence of the large-magnitude Hubbard
 term $\sim U\sum_{i}n_{i\uparrow}n_{i\downarrow}$). Additional features follow from the circumstance that we can "switch-off" completely the Hubbard interaction by applying the magnetic field and saturating magnetically the system. This is possible only (and is the case) because the field induced by the correlations  $h_{corr}$ enhances strongly the effect of applied magnetic field. In that situation a metamagnetic transition takes also place \cite{a31}.

The brief analysis provided above delineates the principal message about what we mean by nonstandard quasiparticles in a (strongly) correlated system. First, they can become quite heavy, i.e. $m_{av}/m_{B}\gg 1$. Second, they depend on the particle-spin direction, what makes them distinguishable in the quantum-mechanical sense, since the mass in nonrelativistic quantum mechanics is an external (input) characteristic in the problem at hand. Third, the effective field driven by the correlations can become very strong i.e. much stronger than any Weiss molecular field appearing in traditional local-moment magnetism. All these microscopic properties must be determined self-consistently. These features of those nonstandard quasiparticles distinguish them from those are defined within the original phenomenological Landau- Fermi-liquid theory, where the enhancement of the effective mass and of the magnetic susceptibility is expressed in terms of interaction parameters. Additionally,
 in the Landau theory of Fermi liquids the enhancement factors are determined by including the interaction only among the quasiparticles at and/or in close vicinity of the Fermi surface. Here, \emph{all the particles} mutually influence each other, what is expressed via an integration over all occupied states when solving appropriate self-consistent equations. This is because in the present situation the interaction is strong (at least comparable) to he Fermi energy.

\subsection*{Superconducting state: Fulde-Ferrell-Larkin-Ovchinnikov state}

We describe next the Fulde-Ferrell-Larkin-Ovchinnikov (FFLO) state for the model of heavy-fermion system starting from Hamiltonian (\ref{D1}) derived in the Appendix D, with the hybrid pairing introduced in the preceding Section.

In the standard BCS approximation with anomalous averages, Hamiltonian (\ref{D1}) in the narrow band limit reduces to the form
\begin{multline}
\mathcal{H} = \sum_{\pmb{k}\sigma}\left(\epsilon_{\pmb{k} \sigma} - \mu\right) f_{\pmb{k} \sigma}^\dagger f_{\pmb{k} \sigma} - \frac{1}{2}g \mu_B H \sum_{\pmb{k}} \left(f_{\pmb{k} \ua}\dg f_{\pmb{k} \ua} - f_{\pmb{k} \da} \dg f_{\pmb{k} \da}\right) \\
+\sum_{\pmb{k}} \left(\Delta_{\pmb{k} \bQ}^* \:f_{\pmb{k} \ua} f_{-{\pmb{k}}+\bQ \da} + H.c.\right) + N \frac{|\Delta_{\pmb{k}\bQ}|^2}{V_0}, \label{H_29}
\end{multline}
with the single-particle energy parametrized in the tight-binding approximation, which in the case of square lattice takes the form
\eqn
\epsilon_{\pmb{k} \sigma} & \equiv & q_\sigma [-2 t (\cos{k_x} + \cos{k_y}) + 4 t' \cos{k_x} \cos{k_y}], \label{E_30}
\eqnx
where $t$ and $t'$ are the first and the second hopping integrals and $q_{\sigma}$ is, as before, the Gutzwiller band narrowing factor (the inverse spin-dependent mass enhancement). The superconducting gap is thus determined from the self-consistent equation
\eqn
\Delta_{\pmb{k} \bQ} & = & -\frac{V_0}{N} \sum_{\pmb{k}'} \gamma_{\pmb{k}} \gamma_{\pmb{k}'} \langle f_{-{\pmb{k}'}+\bQ \da} f_{\pmb{k}' \ua} \rangle.
\eqnx
Additionally, $V_{0}$ is the pairing magnitude which for the constant hybridization has the magnitude
\eq
V_{0}\simeq -\frac{4V^2 \left(q_{\sigma}q_{\bar{\sigma}}\right)^{1/2}}{\epsilon_{f}+U},
\eqx
and the factors $\gamma_{\pmb{k}}$ and $\gamma_{\pmb{k}'}$ correspond to the separable
$\pmb{k}$-dependent factors in (\ref{D3}) divided by $V^{2}$.
\begin{figure}
\begin{center}
\includegraphics[width=0.48\textwidth]{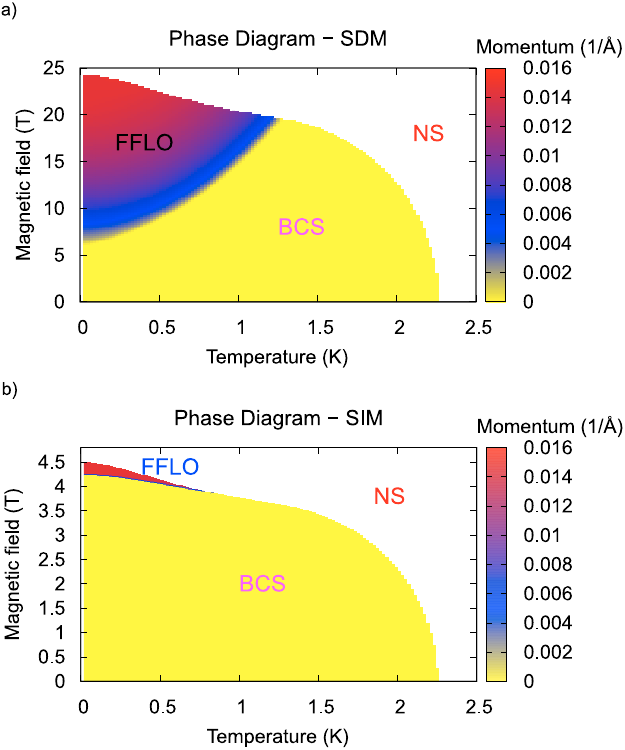}
\caption{(Color online).Phase diagram for the cases with the spin-dependent (a) and the spin-independent masses (b). Yellow region corresponds to $\bQ = 0$ (BCS phase), the darker one to $\bQ \neq 0$ (FFLO phase) and the white to normal state.
Note that with increasing temperature, the transition from BCS to FFLO state occurs at higher fields,
in qualitative agreement with experimental results. The FFLO phase is stable in an extended $H_a$-$T$ regime only in the spin-dependent-masses (SDM) case \cite{a28}.}
\end{center}
\end{figure}

Next, we carry out the approximate form of the Bogoliubov transformation which in the meantime acquired the name of the Bogoliubov-Valatin-de Gennes-Nambu transformation! For that purpose we represent (\ref{H_29}) in the matrix form
\begin{multline}
\mathcal{H} = \sum_{\pmb{k}} (f_{\pmb{k} \ua} \dg, f_{-{\pmb{k}} + \bQ \da} )\times\\
\left( \begin{array}{cc}
\epsilon_{\pmb{k} \ua} - g \mu_B H - \mu & \Delta_{\pmb{k}, \bQ} \\
\Delta_{\pmb{k}, \bQ}^* & -\epsilon_{-{\pmb{k}}+\bQ \da} - g \mu_B H + \mu
\end{array} \right)
\left( \begin{array}{c} f_{\pmb{k} \ua} \\ f_{-{\pmb{k}}+\bQ \da} \dg \end{array} \right)  \\
 + \sum_{\pmb{k}} (\epsilon_{\pmb{k} \da} + g \mu_B H - \mu) + N \frac{\Delta_{\bQ}^2}{V_0}.
\end{multline}
The transformation to the quasiparticle representation has the usual form
\eq
\left( \begin{array}{c} \tilde{\alpha}_{\pmb{k}} \\ \tilde{\beta}_{\pmb{k}} \dg \end{array} \right) = \left( \begin{array}{cc}
u_{\pmb{k}}    &   v_{\pmb{k}} \\
-v_{\pmb{k}}   &   u_{\pmb{k}}
\end{array} \right)
\left( \begin{array}{c} f_{\pmb{k} \ua} \\ f_{-{\pmb{k}}+\bQ \da} \dg \end{array} \right),
\eqx
with the Bogoliubov coherence factors given now by the relations
\eqn
u_{\pmb{k}} = \left[\frac{1}{2}\left(1+\frac{\epsilon_{\pmb{k} \ua} + \epsilon_{-{\pmb{k}}+\bQ \da} - 2 \mu}{\sqrt{(\epsilon_{\pmb{k} \ua} + \epsilon_{-{\pmb{k}}+\bQ \da} - 2 \mu)^2 + 4 \Delta_{\pmb{k} \bQ}^2}}  \right) \right]^{1/2}, \\
v_{\pmb{k}} = \left[\frac{1}{2}\left(1-\frac{\epsilon_{\pmb{k} \ua} + \epsilon_{-{\pmb{k}}+\bQ \da} - 2 \mu}{\sqrt{(\epsilon_{\pmb{k} \ua} + \epsilon_{-{\pmb{k}}+\bQ \da} - 2 \mu)^2 + 4 \Delta_{\pmb{k} \bQ}^2}}  \right) \right]^{1/2}.
\eqnx
The quasiparticle energies in the phase with $\bQ \neq 0$ are
\begin{multline}
E_{\pmb{k}\bQ\alpha}=\frac{1}{2}\left(\epsilon_{\pmb{k} \uparrow}-\epsilon_{-{\pmb{k}}+\bQ \downarrow}\right)-g\mu_{B}H \\
+\alpha\frac{1}{2}\left[\left(\epsilon_{\pmb{k}\uparrow}+\epsilon_{-{\pmb{k}}+\bQ\downarrow} -2\mu\right)^2 +4|\Delta_{\pmb{k}\bQ}|^2\right]^{1/2},
\end{multline}
where sign factor $\alpha = \pm$ corresponds to the electron or hole excitations, respectively.

Having discussed the explicit expression for the fermionic  quasiparticle excitations, we can construct the free energy functional, as well as determine the system of self-consistent equations for $\Delta_{\pmb{k}\bQ}, \mu$, and $\bar{m}$; we also optimize the energy with respect to the magnitude of the wave vector  $\bQ$ \cite{a33}. One should mention that for the
electron-gas situation we have included explicitly in the effective Hamiltonian also the correlation field $h_{corr}$, which we optimize, whereas for the two-dimensional band structure (\ref{E_30}) we have been able so far to carry out only the whole analysis in the Gutzwiller approximation.
In Figs. 9 and 10 we provide the exemplary phase diagrams on the plane temperature $T$
- applied magnetic field $H$ for the three-dimensional gas \cite{a28} and for the square-lattice cases \cite{a33}, respectively. Note that the phase diagrams drawn in the Figures are for the simple Fulde-Ferrell state, i.e. for the form of the gap. Additionally the gap has the $d$-wave symmetry.

\begin{figure}
\begin{center}
\includegraphics[width=0.4\textwidth]{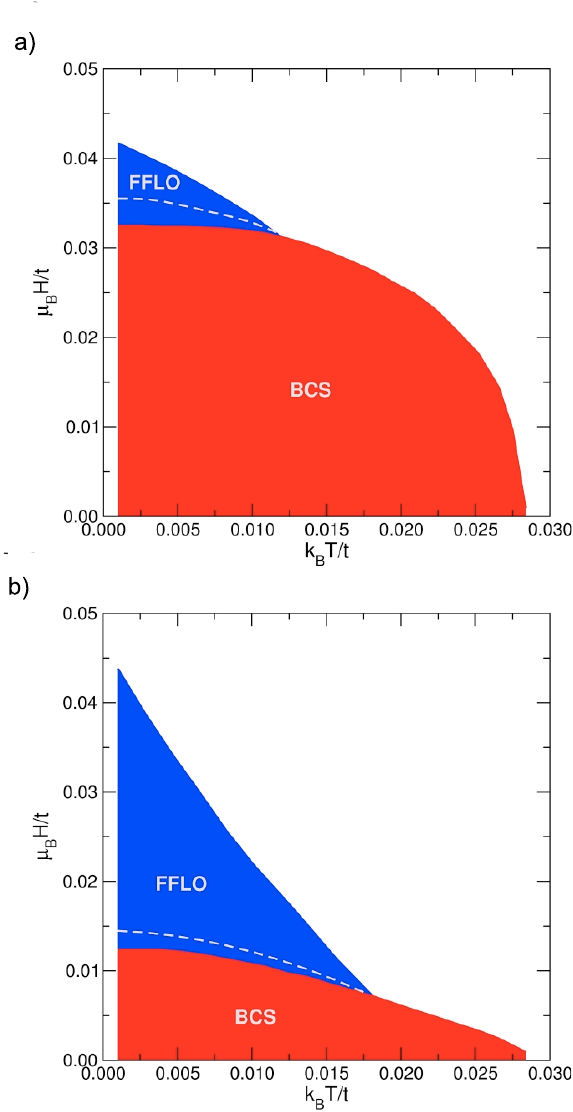}
\caption{(Color online). Phase boundaries for a two-dimensional $d$-wave superconductor with both the spin-independent masses (SIM)
(a) and with the spin-dependent masses (SDM) (b). The FFLO-BCS transition line is of discontinuous nature. The dashed line marks the stability limit of the BCS state as determined by the value of the second critical field $H_{c2}$ for the BCS state. The values of parameters are $n = 0.97$ and $V_0 = 12.5 \, K$. For these values of the parameters, the superconducting transition temperature is $T = 2.5 \, K$ and the uppermost critical field for the FFLO phase is above $6 \, T$. Note that the FFLO state is robust in the situation with SDM and this result is one of the principal features of the present discussion \cite{a33}.}
\end{center}
\end{figure}
\begin{figure}[h!]
\begin{center} 					
{\includegraphics[width=0.45\textwidth]{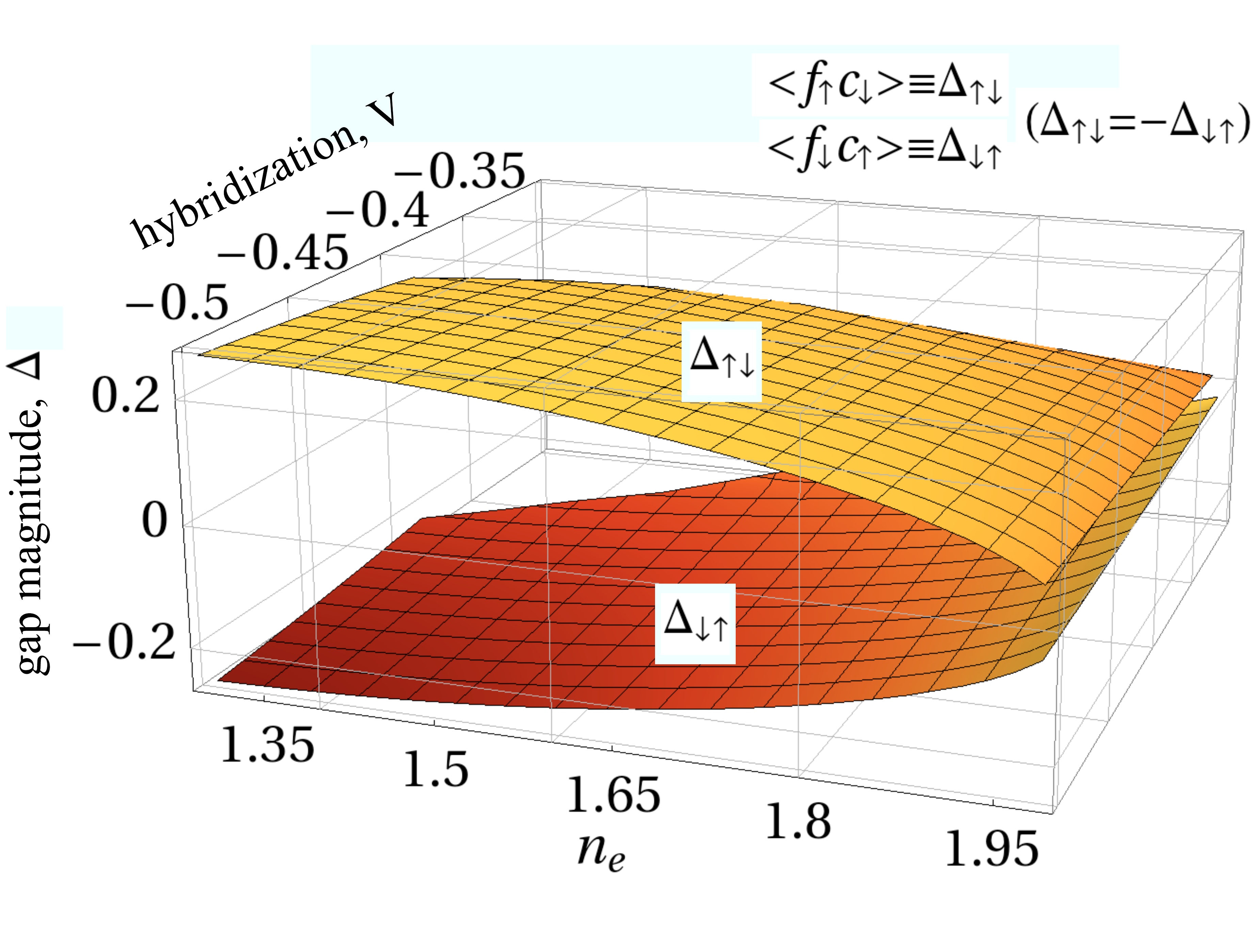}}
\end{center}
\caption{(Color online). Bare unrenormalized hybrid-gap amplitude profiles as a function of both total number $n_{e}$ of electrons per site or intraatomic-hybrid magnitude. The values of parameter $\epsilon_{f}=-1$, $U=3$, $W=2z|t|=1$. The renormalized gap magnitude is $\Delta_{C}\simeq q (\Delta_{\uparrow\downarrow}\Delta_{\downarrow\uparrow})/2$. The gap amplitude vanishes in both Kondo-insulator $(n_{e}\rightarrow 2)$ and in the localized moment ($|V|\rightarrow 0$) limits.}
\end{figure}
The most important feature coming out of these Figures are: (i) the BCS state (i.e. that with $\bQ =0$) is quite robust in the lower  fields and higher temperatures, (ii) the inclusion of the effective-mass spin dependence leads to a remarkable extension of regime of the FFLO stability in both depicted situations, (iii) in the FFLO phase the upper critical field of the transition to the normal phase is much higher than $H_{c2}$ for the BCS superconducting state (cf. Fig. 10, the dashed lines mark the second critical field $H_{c2}$ in the the Pauli limit), and (iv) the first-order BCS$\rightarrow$FFLO transition can be accompanied by a weak metamagnetic transition \cite{a33}. One should also note that the detailed analysis of the FFLO state is carried out separately \cite{a33}. Also, the full SCA analysis of the superconducting states within the full statistically-consistent Fukushima approach for the present model is still to be carried out. We do not expect though, that such
  analysis will change the picture in a decisive manner.

\subsection{Superconducting state by the Kondo-type (hybrid) pairing in the Anderson-Kondo model}

Above, we considered only the narrow $f$-band limit for the Anderson-Kondo model. Such model is valuable if the the $f$-level occupancy $n_{f}$ can be regarded as constant in the considered regime of parameters, i.e. the hybridization gap is regarded as large. Here we mention, that recently we have obtained an explicit solution of the Anderson-Kondo Hamiltonian (\ref{10}) in SCA. The detailed phase diagram will be discussed elsewhere. Here,in Fig. 11 we show the components $\Delta_{\uparrow\downarrow}\equiv\langle f_{i\uparrow}c_{i\downarrow}\rangle$ and $\Delta_{\uparrow\downarrow}\equiv\langle f_{i\downarrow}c_{i\uparrow}\rangle=-\Delta_{\uparrow\downarrow}$, both as a function of magnitude $V$ of the bare (intraatomic) hybridization and the total number of electrons $n_{e}\equiv n_{f}+n_{c}$ (per atomic site containing pair of orbitals $f$ and $c$). The fact, that we treated as separate averages $\langle f_{\uparrow}c_{\downarrow}\rangle$ and $\langle f_{\downarrow}c_{\uparrow}\rangle$ and have obtained that they are of equal magnitude but of opposite sign means, that we have in this case indeed a pure spin-singlet hybrid pairing which vanishes either in the limit of $n_{e}=2$ (where the Kondo-insulator state becomes stable) or when $V\rightarrow 0$, where localized-moment antiferromagnetic phase is stable. Also, the regime of AF-phase stability is separated from that, in which SC is stable.  Work along this lines is in progress and will be reported separately.


\section{Outlook}

Below, instead of making a summary, we pose some important questions concerning the real space pairing first and than conclude by suggesting its universal character applicable also to nuclear and astrophysical quantum matter.

\subsection{From real-space pairing to renormalized-paramagnon mediated pairing}

In this paper we overviewed the concept \emph{of real space pairing}, induced by the kinetic exchange interaction combined with the pair correlated motion. This pairing is facilitated by the circumstance that the kinetic-exchange interaction integral $J$ (or that for the Kondo interaction, $J^{K}$) is comparable to the single-particle energy as expressed by the renormalized hoping magnitude $\sim\!|t|x$ ($|V|x$ for hybridized systems). In that situation, the second exchange-coupled partner to a given electron (or hole) follows in their combined motion throughout the lattice. This situation has been animated graphically elsewhere \cite{a23}. To reiterate, there is no obvious intervening collective boson excitation mediating the pair binding. A contribution of \emph{the renormalized paramagnons} to the pairing represents an additional factor to be evaluated separately starting from RMFT.

Namely, in the situation when the renormalized hopping magnitude can be smaller than the pairing potential magnitude $J$ (cf. Appendix C, when $x\leqslant x_{c1}$), the Hartree-Fock-type (BCS) decoupling of the term, $B_{ij}^{\dagger}B_{ij}\approx \langle B_{ij}^{\dagger}\rangle B_{ij}+ \langle B_{ij}\rangle B_{ij}^{\dagger} - \langle B_{ij}^{\dagger}\rangle\langle B_{ij}\rangle$, can be regarded only as a first-order approximation. \emph{The Renormalized Mean Field Theory} had to be an invented as an approximate treatment of the effective Hamiltonians, albeit not systematic in the field-theoretical sense, that we first perform the saddle-point approximation and include the quantum Gaussian fluctuations next. Those renormalized Gaussian fluctuations should lead to a residual \emph{paramagnon pairing}. The last approach has been outlined briefly in the Appendix E. There, we reiterate more precisely the just mentioned difficulties connected with the division into the real-space
  and renormalized-paramagnon pairing parts. It remains to be seen to what extent the paramagnon part is important. It should be important in the underdoped  regime $x\sim x_{c1}$ as the $|t|x\sim J.$ The division into the mean-field and the paramagnon parts may be easier to carry out in the limit of moderately or weak correlated systems. This is because then the division into the Hartree-Fock and the fluctuation parts is natural.

\subsection{Effective \emph{t-J} and \emph{t-J-V} models should include direct intersite (interorbital) Coulomb interactions?}

There is one additional feature of the models as represented by the starting effective Hamiltonians  (\ref{2}) and (\ref{5}) not discussed in detail. Namely, the respective intersite or interorbital direct Coulomb interaction should be included in the corresponding Hamiltonians. This is because, from one side $J<K_{\langle ij\rangle}$ and $J_{im}^{K}<U_{fc}$, and from the other, such terms are already included in the corresponding full expression of the Dirac spin-exchange operators. Their role many not be so obvious if we decouple them and include anomalous contraction of the type $\langle a_{i\sigma}a_{i\hat{\sigma}}\rangle$. One may think, that when decoupling in the Hartree-Fock manner the term $\sim \nu_{i}\nu_{j}$ (or $\nu_{1}n_{m}$ in the hybrid pairing case), the contribution to the singlet- and the triplet-pairing channels is the same for this spin-independent term and therefore, the whole term can be disregarded when considering the paired state. Additional reason f
 or disregarding the term $\sim K_{ij}$ is that the paring is due to the combined effect of exchange and correlated hopping.
Nonetheless, the inclusion of such terms would present itself as an additional test of the whole picture, though most probably the expected results would be of secondary importance in the most important situations (c.f. also the related discussion of the situation with $K_{ij}<0$ in the main text).

\subsection{From real-space pairing to the Hund's rule exchange pairing}

Once we have suggested, that the exchange interaction is not only a source of spin magnetism
but also a fundamental mechanism of the real-space superconducting pairing, we may ask if
other exchange interactions may become a source of pairing or superfluidity. Here, we have
in view our ideas \cite{a34} about the role of the ferromagnetic, intraatomic, interorbital
(Hund's rule) exchange in orbitally degenerate systems. Customarily, the Hund's rule
exchange is associated with itinerant ferromagnetism in moderately correlated systems
\cite{a35} or with an orbital ordering mixed with magnetic ordering in strongly
correlated systems \cite{a2}. We have suggested that for the itinerant correlated and
orbitally degenerate narrow band systems the Hund's rule exchange provides a stable
spin-triplet superconducting state both in the Hartree-Fock \cite{a35} and in the
strong-correlation limits \cite{a36}. In Fig. 12 we illustrate this statement in the
former limit, respectively by plotting the corresponding phase diagram. The  phase
diagram displayed here supplements the well known phase diagrams involving only magnetic
phases \cite{a36b} with the superconducting paired states. This topics require a detailed
separate analysis, which is in progress.
\begin{figure}   
\begin{center}
\includegraphics[width=0.45\textwidth]{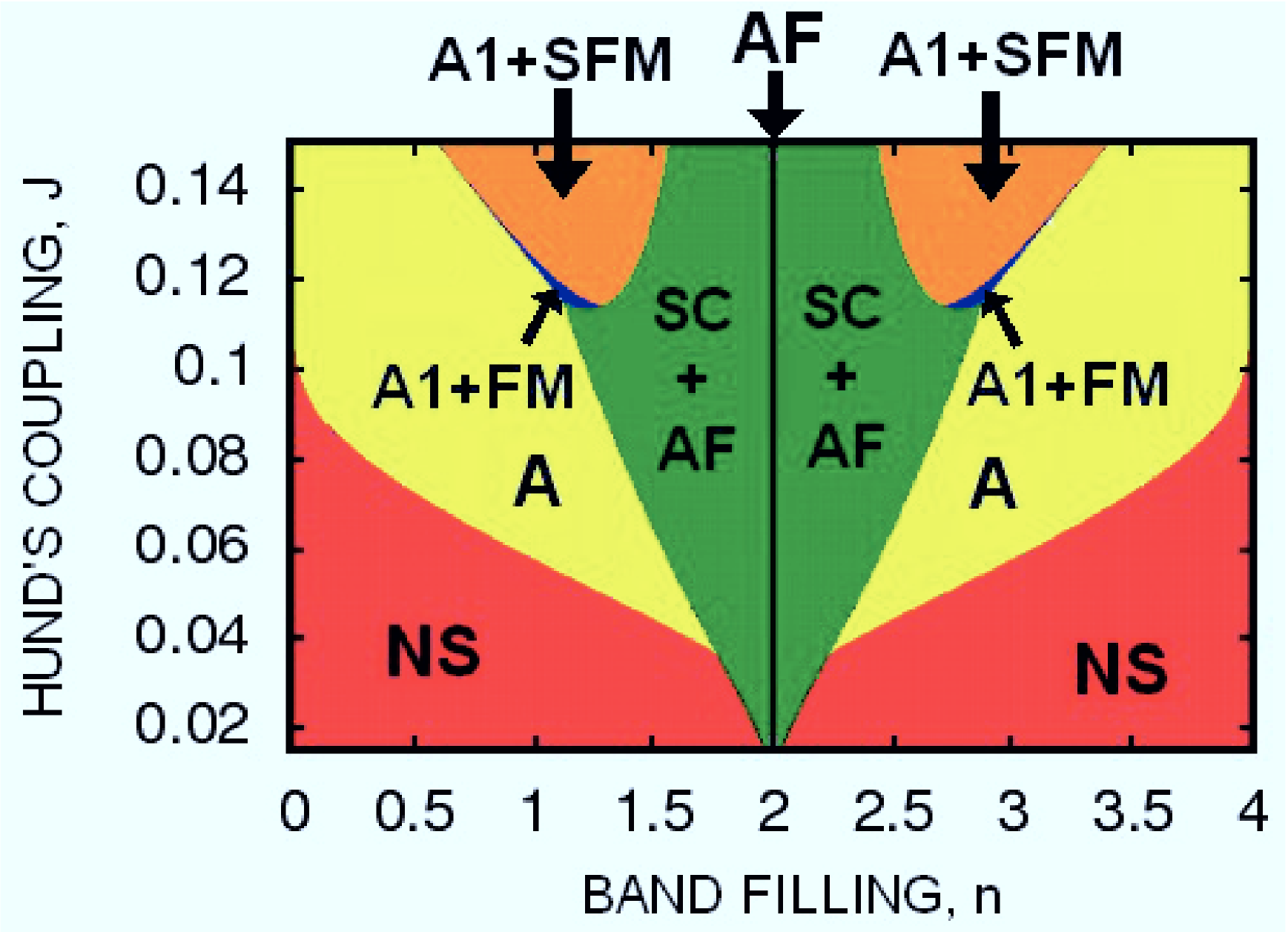}
\end{center}
\caption{(Color online) Exemplary phase diagram \cite{a35} on the plane band filling $n$ - Hund's rule exchange $J=J^{H}$. The phases are as follows: \emph{NS} - normal state; A -superconducting spin-triplet state with the gaps $\Delta_{\uparrow\uparrow}=\Delta_{\downarrow\downarrow}$; $A_{1}$ - superconducting spin-triplet state with the gap $\Delta_{\uparrow\uparrow}\neq 0$ only; \emph{AF} - antiferromagnetic state; \emph{FM} - ferromagnetic state; \emph{SFM} - saturated ferromagnetic state; \emph{AF+SC} - mixed antiferromagnetic + superconducting state with the superconducting gaps different for majority- and minority-spin electrons on given site. The square lattice was assumed with $W=1$, $U=7J$, and no hybridization in the doubly degenerate band was included.}
\end{figure}

\subsection{Concluding remarks}

The real space pairing concept is different from  other concepts, which are based on the idea of virtual boson excitation (phonon, paramagnon etc.) mediating the attraction between fermion. Here the exchange interaction combined with a correlated motion of the pair (pair hopping) is responsible for the pairing. The latter idea has an intuitive interpretation \cite{a23}. However, the real space pairing concepts should be tested further. A selected comparison of RMFT results with experiment have been also discussed in this text. Those results are insufficient, as for instance the Fermi surface evolution and the pseudogap appearance have not been (and cannot be) addressed property within RMFT, i.e. without the phase fluctuations of the superconducting gap.

One additional general remark is in place at the end. Namely, the exchange interaction is a universal interaction in the sense, that it takes place between any two interacting fermions. Therefore, it would be interesting to address the question of real space pairing in a correlated quantum nuclear or astrophysical matter (nucleons in nuclei, neutron stars, quark-gluon plasma, etc.). This topic could become a further example of incorporating the laboratory condensed matter physics into fundamental quantum-matter physics.

\subsection*{Acknowledgments}

This paper overviews very briefly the effort of our TEAM Project awarded by the Foundation for Polish Science (FNP), as well as Grant No. N N202 128736 by the Ministry of Science and Higher Education. I should thank first Dr. Danuta Goc-Jag{\l}o for technical help in assembling this material and the computer animation of real space pairing. I am grateful to my former Ph.D. students: Andrzej Klejnberg, Jakub J\c{e}drak, and Jan Kaczmarczyk for providing detailed analysis of earlier ideas about the exchange-mediated pairing. I am grateful also to my present Ph.D. students: Olga Howczak and Micha{\l} Zegrodnik for extending analysis of exchange-induced pairing to new systems, as mentioned in main text.
Finally, last but not least, I am grateful to my colleagues: Prof. Maciej Ma\'{s}ka, Prof. Tadeusz Doma\'{n}ski, and Prof. Karol Wysoki\'{n}ski for insightful discussions and comments.

\appendix

\section{The concept of $\pmb{t}$-$\pmb{J}$ model and real-space pairing - in perspective}

In 2011 we celebrated 25-th anniversary of the discovery of high-$T_C$ superconductivity. In 2012 there is the 35-th anniversary of our publication of \emph{t-J} model \cite{a38}, as well as 25-th anniversary of discovery its meaning in the context of high-temperature superconductivity. Below I provide a somewhat biased personal account of the last two topics.

\subsection*{1976-1986}

The generalization of the ideas of Anderson \cite{a37,a3} concerning the origin of antiferromagnetic (kinetic) exchange  interaction for the Mott insulators to the strongly correlated metals was was done by the present author in 1976 and subsequently published with his colleagues \cite{a38}. The pioneering period was reviewed in my habilitation-schrift \cite{a39}. The idea of defining the division of the hoppings into the four terms corresponding to the intra- and inter- Hubbard subband hopping was  particulary difficult to envisage. So the resulting canonical transformation, as the "unperturbed" part of the Hamiltonian, was of non-diagonal form in the Fock space. The last feature meant that we had to perform the canonical perturbation expansion in the operator form. A more formal overview  of the pioneering era is given elsewhere \cite{a40}. Many authors have rederived subsequently of the \emph{t-J} model, but I claim that our derivation was the first one, albeit limited to
 the discussion of magnetic phases and the Mott transition as the band filling approaches unity. Some authors claim that Harris and Lange \cite{a41} provided the basis of projected fermion operators $\{a_{i\sigma}^{\dagger}(1-n_{i\hat{\sigma}}),a_{i\sigma}(1-n_{i\hat{\sigma}})\}$ and the kinetic exchange for the $n\neq 1$ case, but a cursory look to the paper shows that it does not contain any explicit mention of the kinetic exchange and does not  introduce $t^{2}/U$ effects explicitly.

One should mention that practically at the same time we have also introduced \cite{a42} an analogical effective Hamiltonian for the Wolff model of the magnetic impurity, which is to a certain degree analogous the Anderson model of magnetic impurity. There, we introduced the cases of "shallow" and "deep" impurity cases, which lead me to the concept of the modified Schrieffer-Wolff transformation in the "shallow-impurity" case. This distinction is elaborated in the main text, when we talk about the hybrid (Kondo) pairing for the Anderson lattice, appearing concomitantly with the itineracy of $f$ electrons.

Practically, nobody was interested in those papers then, as it seemed that the fashionable mixed-valence and heavy-fermion physics had not much to do with the Hubbard model or the Wolff model. There, the periodic Anderson model or the Kondo-lattice models were regarded as the distinct and relevant models.

\subsection*{1987-now}

The revolution in theory of correlated electron systems came with the introduction of real-space pairing amplitudes $\langle a_{i\sigma}^{\dagger}a_{i\bar{\sigma}}^{\dagger}\rangle$ within the \emph{t-J} model. The author learnt about the idea from the preprint of Ruckenstein, Hirschfeld, and Appel \cite{a43}. The often quoted paper \cite{a44} was completely illegible at the time(not only to the author) and it is extremely difficult now to argue, at least for me, who introduced the real space pairing first. We can also say, that we have invented \cite{a6} a correct form of the \emph{t-J} Hamiltonian with the precise real-space projected operators $\{B_{ij}^{\dagger}, B_{ij}\}$, as well as have extended the representation to the case with projected hybrid pairing operators  $\{b_{im}^{\dagger}, b_{im}\}$, which appear in the Anderson-lattice model in "large-but-finite $U$ limit"\cite{a7}. The last model we have termed as \emph{the Anderson-Kondo model}. But, in analogy to \emph{t-J} model, this model should be rather termed \emph{the t-J-V model}, where $V$ stands for hybridization which appears concomitantly with the exchange couplings (Kondo or superexchange) and is instrumental in driving the itineracy of $f$ electrons in heavy-fermion systems.

The second aspect of the real space pairing driven by the kinetic exchange interaction is the absence of the virtual boson driving the real space at least in the mean field approximation. This question leads us to a highly nontrivial problem of formulating renormalized mean field theory (RMFT). This nontriviality of the mean-field-approximation formulation in this situation stems from the fact that the projected hopping part is, strictly speaking, of many-body nature, since it contains $a_{i\sigma}^{\dagger}(1-n_{i\bar{\sigma}})a_{j\sigma}(1-n_{j\bar{\sigma}})$ factor. However, absolutely crucial in the analysis are also the statistically consistency conditions phrased explicitly in our recent works \cite{a5,a13,a14}, as well as in some other papers \cite{a45}. The rationale behind these consistency conditions, expressed through Lagrange multipliers and added to the effective Hamiltonian, is to ensure that the self-consistent equations for the averages appearing in the RMFT p
 rovide the same results as those obtained from an appropriate variational procedure for the free-energy functional representation, the Landau functional in this situation.

The third aspect of the current research is to go beyond the mean-field approximation, i.e. include the Gaussian fluctuations. So far, this approach has been formulated within the slave-boson approach (SBA) \cite{a3}, which however contains spurious phase transitions corresponding to the condensation of auxiliary boson fields. RMFT with the constraints is in some aspects equivalent to SBA in the saddle-point approximation. A systematic approach to incorporate the quantum fluctuations starting from RMFT, is still missing [cf. Appendix E].

An approach based on the Quantum Monte-Carlo method provides a very important insights into the results, albeit limited to very small systems. A combination of RMFT with spatially inhomogenous order parameters and quantum Monte-Carlo methods seem to be also very promising.

\section{Analytic estimate of the BCS type gap magnitude, universal $\pmb{2\Delta(0)/k_{B}T_{C}}$ ratio, and reduction to classic mean field case}

Here we sketch the derivation of Eq. (\ref{32}). Its importance derives from the surprisingly good approach in Fig. 6. of the numerically obtained gap magnitude $\Delta_{C}$ with that from (\ref{32}). Our results put on a solid ground the estimates proposed originally (but not derived) by Rickayzen \cite{a26}.

We start from the self-consistent equation for the gap magnitude within BCS theory, namely
\begin{equation}
1=\frac{1}{2}\int_{-\hbar\omega_{D}}^{\hbar\omega_{D}} d\epsilon\:\rho(\epsilon)V(\epsilon) \frac{1}{\sqrt{\epsilon^{2}+\Delta^{2}}}\tanh{\left(\frac{\sqrt{\epsilon^{2}+\Delta^{2}}} {2k_{B}T} \right)},
\label{B1}
\end{equation}
where $V(\epsilon)$ is the absolute value of the pairing potential, $\rho(\epsilon)$ is the density of states in the band, $\epsilon$ is the particle energy counted from the Fermi energy $\epsilon_{F}\equiv 0$, and $\hbar\omega_{D}$ is the energy cut-off for the pairing potential (in the real-space pairing case, we take $\hbar\omega_{D}\simeq J$, the magnitude of kinetic-exchange interaction). We believe, it is reasonable to make the so-called BCS approximation at this stage, justified in the present situation in the following manner. Namely, we utilize Rolles's theorem about the average value of the defined integral:
\begin{equation}
\int_{a}^{b}dx\:f(x)=f(\bar{x})(b-a),
\label{B2}
\end{equation}
 where $\bar{x}\in [a,b]$. In other words, the area under the curve $f(x)$ can be represented by that of a rectangular. In applying this theorem to (\ref{B1}) we find that it can be rewritten in the form:
 \begin{equation}
 1=\hbar\omega_{D}\rho(\bar{\epsilon})\frac{1}{\sqrt{\bar{\epsilon}^{2}+\Delta^{2}}} \tanh{\left(\frac{\sqrt{\bar{\epsilon}^{2}+\Delta^{2}}}{2k_{B}T}\right)}.
 \label{B3}
 \end{equation}
Now, it is reasonable to assume that $\Delta\ll\hbar\omega_{D}\ll\epsilon_{F}$, so we can represent $\epsilon_{k}-\mu$ in the same manner, as in the Fermi liquid theory, i.e., assume that near the Fermi level
\begin{equation}
\epsilon_{\pmb{k}}-\mu\equiv\hbar v_{F}k,
\label{B4}
\end{equation}
so that in volume $\Omega$ the density of states is
\begin{equation}
 \rho(\epsilon)=\frac{\Omega}{\left(\hbar v_{F}\right)^{3}\pi^{2}}\;\epsilon^{3},\;\;\;
 \mbox{with}\;\;\; \epsilon=\hbar v_{F}k,
\label{B5}
\end{equation}
which is an antisymmetric function of $\epsilon$. Hence, the integral of (\ref{B2}) is an increasing function as we cross the Fermi level energy $\mu$. This minor argument is used to make a bold statement, that in such a situation an almost exact approximation can be made , i.e. one can put  $\bar{\epsilon}=0$ in (\ref{B3}). This means that the integral equation can be replaced by the following algebraic equation
\begin{equation}
1=\hbar\omega_{D}V(0)\rho(0)\frac{1}{\Delta(T)}\tanh{\left[\frac{\Delta(T)}{2k_{B}T}\right]},
\label{B6}
\end{equation}
which represents a classical mean-field self-consistent equation for the order parameter $\Delta(T)$ (e.g. for the spin $(1/2)$ Ising model). Parenthetically, this means that the BCS-superconductor phase transition is of classical nature, with a mean-field (van-der-Waals type) classical critical point $T_{C}$.

One can draw additional conclusions from the result (\ref{B6}). First, assuming that the order parameter $\Delta=\Delta(T)$ is small when $T\rightarrow T_{C}-0$, one can rewrite (\ref{B6}) in the form of Landau equation for the order parameter:
\begin{equation}
\Delta\simeq\hbar\omega_{D}V(0)\rho(0)\left[1-\frac{1}{3}\left(\frac{\Delta} {2k_{B}T}\right)^{2}\right] \frac{\Delta}{2k_{B}T}
\label{B7}
\end{equation}
or equivalent (for $T\simeq T_{C}$):
\begin{equation}
\Delta\left[1-\frac{\hbar\omega_{D}V\rho}{2k_{B}T}\right]+\frac{1}{24} \frac{\hbar\omega_{D}V\rho}{\left(k_{B}T_{C}\right)^{3}}\;\Delta^{2}=0.
\label{B8}
\end{equation}
Thus, to a good approximation
\begin{equation}
k_{B}T_{C}=\frac{1}{2}\omega_{D}V(0)\rho(0)
\label{B9}
\end{equation}
and then it can be rewritten in the form
\begin{equation}
\frac{\Delta(T)}{2k_{B}T_{C}}=\tanh{\left[\frac{\Delta(T)}{2k_{B}T}\right]}.
\label{B10}
\end{equation}
This relation, in turn, leads to the following universal relation for the gap-to-$T_{C}$ ratio:
\begin{equation}
\frac{2\Delta(0)}{k_{B}T_{C}}=4.
\label{B11}
\end{equation}
This value differs from the usual estimate of this ratio - the value 3.53 (or 3.311) - obtained directly from Eq. (\ref{B1}) estimate. We think the discrepancy is not large , in view of even larger difference in experimental data for the BCS superconductors. In the context of our present discussion of correlated systems, this ratio of maximal amplitude $\Delta_{c}(0)$ to the real experimental value of $T_{C}$ should be either equal to the (\ref{B11})  or even larger. The latter situation takes place if the pairing-potential renormalization is augmented by the additional renormalization of the density of states, so that $2\Delta(0)/T_{C}=4(q^{t})^{-1}$.

On the basic of the above treatment one can also rewrite equation (\ref{B6}) in the form
\begin{equation}
\frac{\Delta(T)}{\Delta(0)}=\tanh{\left[\frac{\Delta(T)}{\Delta(0)}\;\frac{T_{C}}{T}\right]},
\label{B12}
\end{equation}
which coincides with that proposed originally by Rickayzen \cite{a26}. In brief, this discussion, ignited by the coincidence of the results shown in Fig. 6, illuminates a "BCS aspect" of the present renormalized mean field theory.

\section{Elementary estimate of the lower critical concentration for onset of antiferromagnetism}

Here we estimate the critical concentration for the antiferromagnetism disappearance as a function of doping. For that purpose we start from the canonical version of the \emph{t-J} model (\ref{2}) without the last two terms included. One may say that in the Mott insulating state antiferromagnetic interaction dominates over the hole hopping and this means that the (lower) critical concentration is achieved when the two contributions are of the same magnitude, i.e.,
\begin{equation}
-z\:|t|\:x(1-x)+J(1-x)^{2}=0.
\label{C1}
\end{equation}
This happens roughly at the hole concentration
\begin{equation}
 x=x_{c_{1}}\simeq \frac{J}{z\:|t|\:+J}\simeq \frac{1}{z}\;\frac{J}{|t|},
\label{C2}
\end{equation}
 which yields $x_{c_{1}}\simeq 0.07$, a quite reasonable value in view of the simplicity of the estimation.

In order to try to improve the estimate we formulate an elementary Gutzwiller type of approach in which the renormalized hopping and the exchange parts are of the same amplitude. This means that we can postulate the ground energy per site of the form
\begin{equation}
\frac{E_{G}}{N}=-z\:|t|\:\Phi(\lambda)\sum_{j(i)}\langle a_{i\sigma}^{\dagger}a_{j\sigma}\rangle_{0}-Jz\lambda n^{2},
\label{C3}
\end{equation}
where the variational parameter $\lambda\equiv \langle(1/4)n_{i}n_{j}-\pmb{S}_{i}\cdot\pmb{S}_{j}\rangle$, is equal to unity in the N\'{e}el state of the Mott insulator and reduces to the value $\lambda=(1/4)\langle n_{i}\rangle \langle n_{j}\rangle=1/4$ in the uncorrelated state. The renormalization factor can be expanded in powers of $\lambda$
\begin{equation}
\Phi(\lambda)=g_{0}+g_{1}\lambda+g_{2}\lambda^{2},
\label{C4}
\end{equation}
in which higher order terms have been disregarded as we are close to the magnetism disappearance (or onset). Also, the hopping probability in the uncorrelated state is $\langle a_{i\sigma}^{\dagger}a_{j\sigma}\rangle_{0}=n_{\sigma}(1-n_{\sigma})=(n/2)(1-n/2)$. Next, the constant $g_{0}$, $g_{1}$, and $g_{2}$ can be determined from the particular solvable limits, namely:
\begin{eqnarray}
 &\mbox{1}^{\mbox{\rm o}} \;\;\;\;\;  \Phi(\lambda=1)=0 \label{C5} \\
 &  \mbox{we do not encounter any hopping,  i.e.,}\;\;\; g_{0}+g_{1}+g_{2}=0
\nonumber\\
& \mbox{2}^{\mbox{\rm o}} \;\;\;\;\;  \Phi(\lambda=\frac{1}{4})=1\label{C6}\\
& \mbox{i.e.,}\;\;\; g_{0}+\frac{1}{4}g_{1}+\frac{1}{16}g_{2}=1.
 \nonumber
\end{eqnarray}
Hence, the optimal value of $\lambda=1/4$ (from $\partial{E_{0}}/\partial{\lambda}=0$) becomes
\begin{equation}
\lambda=\frac{1}{4g_{2}}\;\frac{Jn}{|t|\:(1-n)}+g_{1}.
\label{C7}
\end{equation}
Now, the remaining condition is that $\lambda=1/4$ when $J=0$. In result, we obtain $g_{0}=-\frac{1}{9}$, $g_{1}=\frac{8}{9}$, and $g_{2}=-\frac{16}{9}$. Hence,
\begin{equation}
x_{c_{1}}=1-n_{c_{1}}\simeq\frac{1}{1+\frac{8\:|t|}{3J}}\simeq0.1,
\label{C8}
\end{equation}
for $|t|/J=3$. Finally, the optional ground state energy per pair of states and for $n\geqslant n_{c_{1}}$ is
\begin{equation}
\frac{E_{G}}{zN}=-|t|\:n(1-n)-\frac{J}{4}n^{2}-\frac{9}{64}\:\frac{J^{2}n^{3}}{|t|\:(1-n)}.
\label{C9}
\end{equation}
Note that if the ratio $|t|/J=6$, then $x_c{_{1}}\sim0.07$. This is still too high value by a factor of 2 when compared to the experimental value: $x_{c_{1}}\simeq 0.03$ for $La_{2-x}Sr_{x}CuO_{4}$. Nevertheless, the expression (\ref{C8}) has a straightforward interpretation. Namely, the first term is the renormalized band energy in the Gutzwiller approximation in the normal state and in the $U=\infty\; (J=0)$ limit. The second term is the energy of the N\'{e}el state in the mean field approximation for the Heisenberg part. The last contribution is the higher-order contribution coming from the correlations (note that our order parameter $\lambda$ in the Landau expansion (\ref{C3}) is of two-particle nature, but still we recover a single-particle (mean-field) description).

This lower concentration $n_{c_{1}}$ (or $x_{c_{1}}$) is missing in the present approach (cf. Sec. IV C), since no localization effects within RMFT-SCA were taken into consideration. This means that we may need to supplement our whole approach with some sort of cluster expansion, i.e. improve on the Gutzwiller (or Fukushima) ansatz. We should see a progress along this line in near feature e.g. by evaluating the averages with the help of a full Gutzwiller wave function \cite{a22}.

\section{From Kondo coupling to real space pairing in an almost localized Fermi liquid of heavy quasiparticles} \label{appD}

Hamiltonian (\ref{10}) can be transformed to the mean-field form and subsequently to the momentum representation, in which it takes the form \cite{a33}
\begin{multline}
\mathcal{H} = \sum_{\pmb{k} \sigma} \left[ \epsilon_{\pmb{k} \sigma} c_{\pmb{k} \sigma}\dg c_{\pmb{k} \sigma} + \widetilde{\epsilon}_{f\sigma} f_{\pmb{k}\sigma}\dg f_{\pmb{k} \sigma} + \widetilde{V}_{\pmb{k} \sigma} f_{\pmb{k} \sigma}\dg c_{\pmb{k} \sigma} \right.\\
\left. + \widetilde{V}_{\pmb{k} \sigma}^* c_{\pmb{k} \sigma}\dg f_{\pmb{k} \sigma} \right]
- \frac{2}{\ef+U} \frac{1}{N} \sum_{\pmb{k} \pmb{k}' \mathbf{Q}} V_{\pmb{k}} V_{\pmb{k}'}^* \left(q_{\sigma}q_{\bar{\sigma}}\right)^{1/2} A_{\pmb{k}, \mathbf{Q}}\dg A_{\pmb{k}', \mathbf{Q}}, \label{D1}
\end{multline}

with

\eq
A_{\pmb{k} \mathbf{Q}}\dg = \frac{1}{\sqrt{2}} \Big( f_{\pmb{k} + \mathbf{Q}/2 \ua}\dg c_{-\pmb{k} + \mathbf{Q}/2 \da}\dg - f_{\pmb{k} + \mathbf{Q}/2 \da}\dg c_{-\pmb{k} + \mathbf{Q}/2 \ua}\dg \Big). \label{D2}
\eqx
The wave vector $\pmb{Q}$ is nonzero in FFLO phase.
The single-particle part can be easily diagonalized by moving to the hybridized basis, whereas the pairing part is represented by a \textit{separable} pairing potential. We shall proceed with the transformation to the hybridized basis first, which yields the following transformed pairing part for the lower hybridized band
\begin{multline}
\mathcal{H} = \sum_{\pmb{k} \sigma} E_{\pmb{k} \sigma} \alpha_{\pmb{k} \sigma}\dg \alpha_{\pmb{k} \sigma}
 - \frac{4}{\ef+U}\sum_{\pmb{k} \pmb{k}' \mathbf{Q}} \\
 \frac{|V_{\pmb{k}} V_{\pmb{k}'}|^2 (q_\sigma q_{\os})^{1/2}}{[(\epsilon_{\pmb{k}\sigma} - \tilde{\epsilon}_{f\sigma})^2 + |\tilde{V}_{\pmb{k}\sigma}|^2]^{1/2} [(\epsilon_{\pmb{k}'\os} - \tilde{\epsilon}_{f\os})^2 + |\tilde{V}_{\pmb{k}'\os}|^2]^{1/2}} \times \\
 \times \alpha_{\pmb{k} + \mathbf{Q}/2 \ua}\dg \alpha_{-\pmb{k} + \mathbf{Q}/2 \da}\dg \alpha_{-\pmb{k}' + \mathbf{Q}/2 \da} \alpha_{\pmb{k}' + \mathbf{Q}/2 \ua},
 \label{D3}
\end{multline}
with the dispersion relation in the lower hybridized band in the form
\eq
E_{\pmb{k} \sigma} \equiv \frac{\epsilon_{\pmb{k} \sigma} + \tilde{\epsilon}_{f\sigma}}{2} - \Big[ \Big(\frac{\epsilon_{\pmb{k} \sigma} + \tilde{\epsilon}_{f\sigma}}{2}\Big)^2 + |\tilde{V}_{\pmb{k}\sigma}|^2 \Big]^{1/2}. \label{D4}
\eqx
Note that in (\ref{D3}) and (\ref{D4}) we have written the formulas for the spin-polarized situation. For real $V_{\pmb{k}}$, the hybridized quasiparticle operator reads $\alpha_{\pmb{k} \sigma} = \cos{\theta_{\pmb{k} \sigma}} f_{\pmb{k} \sigma} + \sin{\theta_{\pmb{k} \sigma}} c_{\pmb{k} \sigma}$, with the condition for the mixing angle $\theta_{k}$
\eq
\tan{2 \theta_{\pmb{k} \sigma}} = \frac{2 \tilde{V}_{\pmb{k} \sigma}}{\epsilon_{\pmb{k} \sigma} - \tilde{\epsilon}_{f\sigma}}.
\eqx
Taking the states on the Fermi surface we have $\tan{2 \theta_{\pmb{k} \sigma}} \approx 2
\theta_{\pmb{k} \sigma}$, and hence $\theta_{\pmb{k} \sigma} = \tilde{V}_{\pmb{k} \sigma} / \tilde{\epsilon}_{f \sigma}$. In effect, $\alpha_{\pmb{k} \sigma} \simeq f_{\pmb{k} \sigma}$. The last estimate provides us with the starting Hamiltonian (\ref{H_29}) in the single-band limit. The complicated $\pmb{k}$-dependence of the pairing potential means that in general, the nature of the superconducting gap may take a form more complicated than pure extended $s$-wave or $d$-wave forms. Note also that even though the pairing potential contains explicitly the spin quantum numbers, it is in fact spin independent, as assumed in the main text.

\section{Incorporation of quantum fluctuations within the SGA - RMFT approach and the effective fermion-boson model} \label{appe}

The effective Hamiltonian (\ref{18}) in SGA approximation can be generalized to include the quantum fluctuations. Namely, we can generalize it the spin rotationally invariant form, which can be now written in the following real-space form:
\begin{equation}
\widetilde{\mathcal{H}}\approx \mathcal{H}_{GA}-\sum_{i}\pmb{\lambda}_{i}^{(m)}\cdot \left( \pmb{S}_{i}-\langle \pmb{S}_{i}\rangle \right) - \sum_{i\sigma}\lambda_{i}^{(n)} \left(n_{i\sigma}-\langle n_{i\sigma}\rangle\right),
\end{equation}
where
\begin{equation}
\pmb{S}_{i}\equiv \frac{1}{2}\sum_{\sigma\sigma^{'}}(\pmb{\tau})_{\sigma\sigma^{'}} a_{i\sigma}^{\dagger}a_{i\sigma^{'}}
\end{equation}
and $\pmb{\tau}\equiv (\tau_{x},\;\tau_{},\;\tau_{z})$ are the Pauli matrices. The constraints can be divided into the mean-field and the fluctuation parts according to
\begin{eqnarray}
\widetilde{\mathcal{H}}\approx \mathcal{H}_{GA}&-&\pmb{\lambda}^{(m)}\cdot\sum_{i} \left( \pmb{S}_{i}-\langle \pmb{S}_{i}\rangle \right)\nonumber \\
&-&\lambda^{(n)}\sum_{i\sigma} \left( n_{i\sigma}-\langle n_{i\sigma}\rangle \right) \nonumber\\
&-&\sum_{i}\left(\pmb{\lambda}_{i}-\pmb{\lambda}^{(m)}\right) \cdot \left(\pmb{S}_{i}-\langle\pmb{S}_{i}\rangle\right)\nonumber\\
&-&\sum_{i}\left(\lambda_{i}^{(n)}-\lambda^{(n)}\right) \left(n_{i\sigma}-\langle n_{i\sigma}\rangle\right),
\end{eqnarray}
where we have assumed that $\langle\pmb{S}_{i}\rangle$ is oriented along $z$-axis, if $\pmb{\lambda}^{(m)}\equiv (0,0,\lambda^{(m)})$ is taken. The first three terms compose the SGA Hamiltonian (\ref{18}) (the factor $1/2$ in the second term is irrelevant, since we can redefine $\lambda^{(m)}\equiv\frac{1}{2}\lambda^{(m)}$), whereas the remaining two terms represent the fluctuation around the mean-field equilibrium. We can write down the fluctuation part in the form
\begin{equation}
\delta\widetilde{H}\equiv-\sum_{i}\left(\delta\pmb{\lambda}_{i}^{(m)}\cdot \delta\pmb{S}_{i}+\delta\lambda_{i}^{(n)}\delta n_{i}\right).
\label{E4}
\end{equation}

It is composed of the local spin and the charge fluctuations coupled to the fluctuating Bose fields $\delta\pmb{\lambda}_{i}^{(m)}$ and $\delta\pmb{\lambda}_{i}^{(n)}$. Note that we have neglected the fluctuations in double-occupancy $d_{i}^{2}=\langle n_{i\uparrow}n_{i\downarrow}\rangle$, which can also be included, at least in principle. The form (\ref{E4}) represents the starting point of analysis, which will not be detailed here.

\subsection*{From \emph{t-J} to the fermion-boson model with correlations: A formal transformation}

The above intuitive picture can be rephrased in a more formal language. We outline this method for the t-J model containing explicitly the real space pairing. Namely, we start from the expression for the partition function of \emph{t-J} model (for the simplest situation, i.e., with no three-site terms and with $K_{ij}\equiv 0$) in the coherent-state representation, which reads
\begin{multline}
Z=\int D\left[b_{i\sigma},b_{i\sigma}^{\dagger}\right]\\
\exp\left\{-\int_{0}^{\beta}d\tau\:\sum_{ij\sigma}
\left[ b_{i\sigma}^{\dagger}(\tau)\left[\left(\partial/\partial_{\tau}-\mu\right)\delta_{ij} +t_{ij}\right] \right.\right.\\
\left. \vphantom{\int_{0}^{\beta}} \left.\vphantom{b_{i\sigma}^{\dagger}}  b_{j\sigma}(\tau)+H_{I}(\tau)\right]\right\},
\end{multline}
where we integrate over the Grassmann fields $(b_{i\sigma}^{\dagger},b_{j\sigma})$, $\beta=(k_{B}T)^{-1}$ is the inverse temperature, and
\begin{equation}
\mathcal{H}_{I}(\tau)=-\sum_{\langle ij\rangle}\!^{'} J_{ij}B_{ij}^{\dagger}(\tau)B_{ij}(\tau)
\end{equation}
is the interaction part.
Now, the novel basic idea is that the summation over bonds $\langle ij\rangle$ can be regarded formally as a simple summation; $B_{ij}^{\dagger}B_{ij}$ then has the form appropriate for the Hubbard-Stratonovich transformation, i.e., we can linearize the quartic term in the following manner:
\begin{multline}
\mathcal{H}_{I}\equiv -g\sum_{\langle ij\rangle} B_{ij}^{\dagger}B_{ij}\\
\rightarrow\;\sum_{\langle ij\rangle} \left\{-B_{ij}^{\dagger}\Delta_{ij}-\Delta_{ij}^{*}B_{ij}+\frac{\Delta_{ij}\Delta_{ij}^{*}} {g} \right\},
\end{multline}
with $g\equiv J_{\langle ij\rangle}=J$. In effect, the partition function takes the form
\begin{multline}
Z=\int D\left[b_{i\sigma},b_{i\sigma}^{\dagger},\Delta_{ij},\Delta_{ij}^{*} \right]\\
*\exp\left\{-\int_{0}^{\beta}d\tau \left\{\sum\!^{'}_{ij\sigma} b_{i\sigma}^{\dagger}(\tau) \left[\left(\partial/\partial\tau-\mu\right)\delta_{ij}+t_{ij}\right]b_{j\sigma}(\tau) \right.\right.\\
\left.- \sum_{\langle ij\rangle}\left[\Delta_{ij}(\tau)B_{ij}^{\dagger}(\tau)+ \Delta_{ij}^{*}(\tau)B_{ij}(\tau)-\frac{|\Delta_{ij}|^{2}}{J_{ij}}\right]\right\}\label{E8}
\end{multline}
In effect, we have the following Hamiltonian when we include both fields:
\begin{eqnarray}
\widetilde{\mathcal{H}}\equiv \widetilde{\mathcal{H}}_{FB}(\tau)&=&\sum_{ij\sigma}\!^{'}\left(t_{ij}-\mu\delta_{ij} \right) b_{i\sigma}^{\dagger}(\tau)b_{ij\sigma}(\tau) \nonumber\\
&-&\sum_{\langle ij\rangle}\left[\Delta_{ij}(\tau) B_{ij}^{\dagger}(\tau)+\Delta_{ij}^{*}B_{ij}(\tau)\right] \nonumber\\
&+&\sum_{\langle ij\rangle} J_{ij}^{-1}\Delta_{ij}^{*}(\tau)\Delta_{ij}(\tau).
\label{E9}
\end{eqnarray}
We see that $\Delta_{ij}(\tau)$ in the RMFT is played by $\Delta_{ij}=J_{ij}\langle B_{ij}\rangle$. Here, the representation is exact and the "time $\tau$" dependent pair-fermion field  $B_{ij}^{\dagger}(\tau)$ is coupled directly to the bosonic field $\Delta_{ij}(\tau)$ \cite{a47}, the latter is a Gaussian fluctuating complex field. Alternatively, one may add the Bose term $-\sum_{\langle ij\rangle} J_{ij}^{-1}\Delta_{ij}^{*}\Delta_{ij}$ to the effective Hamiltonian as in (\ref{E9}). In the saddle-point approximation the $\tau$ dependence of the fields is disregarded and the pairing amplitudes are minimized for the free-energy functional. Note, that $\Delta_{ij}$ represent the spin-singlet-pairing amplitudes, as $B_{ij}^{\dagger}$ and $B_{ij}$ have the form (\ref{4}). However, even in the mean-field approximation, defined in the above manner, the problem is still highly nontrivial, since the first term in (\ref{E9}) contains the projected fermionic operators $b_{i\sigma}^{\d
 agger}$ and $b_{j\sigma}$ with non-fermionic anticommutation relations (\ref{3}). This was actually the reason for the discussion on  nontriviality of the renormalized mean-field theory (RMFT). This circumstance leads also to the conclusion, that if we start our analysis from RMFT for (\ref{E9}), then to be able to include the fluctuations in $\Delta_{ij}$ in a systematic manner, one has to include the fluctuations in projected fermionic fields around RMFT solution as well, which may represent a formidable task. Most probably, the viable method to diagonalize exactly (\ref{E9}) in the saddle-point approximation is to resort to the Quantum Monte-Carlo or related cluster-expansion methods. Nonetheless, the expression (\ref{E8}) provides a hint concerning the direction, in which a systematic approach towards rigorous solution of the \emph{t-J} model should be developed.


\begin{thebibliography}{199}
\bibitem{a16} The same conerns the Bardeen-Cooper-Schieffer (BCS) theory where one takes as gap $\Delta_{\pmb{k}}\equiv \left\langle c_{\pmb{k}\uparrow}^{\dagger}\; c_{-\pmb{k}\downarrow}^{\dagger}\right\rangle$. The spin singlet nature of the state is implicitely assumed in the choice of the BCS wave function: $|\Psi\rangle_{BCS}=\prod_{\pmb{k}}(u_{\pmb{k}}+v_{\pmb{k}}c_{\pmb{k}\uparrow}^{\dagger} c_{-\pmb{k}\downarrow}^{\dagger})|0\rangle$, instead: $|\Psi\rangle_{BCS}=\prod_{\pmb{k}}[u_{\pmb{k}}+\frac{1}{2}v_{\pmb{k}} (c_{\pmb{k}\uparrow}^{\dagger} c_{-\pmb{k}\downarrow}^{\dagger}- c_{\pmb{k}\downarrow}^{\dagger} c_{-\pmb{k}\uparrow}^{\dagger})]|0\rangle$ for the spin singlet state $c_{\pmb{k}\uparrow}^{\dagger}c_{-\pmb{k}\downarrow}^{\dagger} =-c_{\pmb{k}\downarrow}^{\dagger}c_{-\pmb{k}\uparrow}^{\dagger}$.
\end{thebibliography}
\end{document}